\documentclass[aps,prl,preprint,superscriptaddress]{revtex4-2}

\usepackage[utf8]{inputenc}
\usepackage{amsmath}
\usepackage{fixmath}
\usepackage{accents}
\usepackage{graphicx}%

\usepackage{hyperref}
\usepackage{physics}
\usepackage{comment}

\newcommand{\re}[1]{\operatorname{Re}\left[#1\right]}
\usepackage{amssymb}

\newcommand{\mb}{\mathbold}

\begin{document}

\title{Efficient generation of entangled photons in the telecommunications range using nonlinear metasurfaces integrated with ScAlN/GaN heterostructures}

\author{Jaeyeon Yu}
\affiliation{Center for Integrated Nanotechnologies, Sandia National Laboratories, Albuquerque, New Mexico, USA}

\author{Jewel Mohajan}
\affiliation{Princeton University, Princeton, NJ, USA}

\author{Mikhail Tokman}
\affiliation{Ariel University, Ariel, Israel}

\author{Jackson Stewart}
\affiliation{Texas A\&M University, College Station, TX, 77843 USA}

\author{Anthony Rice}
\affiliation{Sandia National Laboratories, Albuquerque, New Mexico, USA}

\author{Sadhvikas Addamane}
\affiliation{Center for Integrated Nanotechnologies, Sandia National Laboratories, Albuquerque, New Mexico, USA}

\author{Oana Malis}
\affiliation{Purdue University, West Lafayette, IN, USA}

\author{Alejandro W. Rodriguez}
\affiliation{Princeton University, Princeton, NJ, USA}

\author{Igal Brener}
\affiliation{Center for Integrated Nanotechnologies, Sandia National Laboratories, Albuquerque, New Mexico, USA}

\author{Raktim Sarma*}
\affiliation{Center for Integrated Nanotechnologies, Sandia National Laboratories, Albuquerque, New Mexico, USA}

\author{Alexey Belyanin}
\affiliation{Texas A\&M University, College Station, TX, 77843 USA}

\date{\today}

\begin{abstract} 

Entangled photons provide non-classical correlations that enable measurement sensitivities beyond classical limits, scalable fault-tolerant quantum computation, and fundamentally secure quantum communication, making them a foundational necessity for next-generation quantum technologies. Here we propose and analyze a novel source of entangled photons based on ScAlN/GaN quantum wells integrated with dielectric metasurfaces. Giant second-order intersubband nonlinearity of the GaN quantum wells with strain-compensated delta-doped ScAlN barriers caused by strong built-in electric fields combined with superior mode-coupling performance of metasurfaces optimized by inverse design give rise to efficient parametric down-conversion and generation of entangled photons in the telecom range. We develop a rigorous Heisenberg-Langevin formalism which includes field quantization, dissipation and fluctuations for all fields, parametric amplification of thermal noise and zero-point fluctuations, and other relevant effects. Our proposed approach of employing the emergent photonic material ScAlN promises high biphoton generation rate over $10^{10}$ s$^{-1}$ from a compact integrated structure  that is only 0.5 $\mu$m thick while mitigating strain-related issues that have so far impeded progress of nitride-based heterostructures for quantum photonic applications into the infrared and visible wavelengths. Our result therefore is relevant for numerous applications ranging from quantum sensing, quantum information, and computing. 

\end{abstract}

\maketitle

\noindent
\textbf{\#Corresponding author:} Raktim Sarma (rsarma@sandia.gov)\\
\noindent
\textbf{Equal contribution:} Jaeyeon Yu and Jewel Mohajan contributed equally to this work.

\section{Introduction}

Highly nonlinear materials and nanostructures capable of high-efficiency generation of entangled photon pairs and heralded single photons via spontaneous parametric down-conversion (SPDC) process are expected to become a crucial part of future integrated quantum photonic circuits \cite{solntsev2017, wang2020, sergei, ma2020, moody2020,tokman2022a,tokman2023a}. 
Semiconductor quantum-well (QW) heterostructures are among the best candidates for such ultracompact, highly nonlinear materials. They are known to possess giant resonant second-order optical nonlinearity associated with intersubband transitions when the confining potential breaks the inversion symmetry \cite{sirtori1991,gmachl2006}. The latter is usually achieved by using step or coupled asymmetric QWs. When integrated with quantum cascade lasers, intersubband nonlinearity has enabled efficient frequency up-conversion and difference frequency generation \cite{malis2004,belkin2007} leading to room-temperature operated THz laser sources \cite{belkin2008}. 

However, for many applications it is desirable to have passive nonlinear QW nanostructures illuminated by normally incident light. In this case the use of intersubband nonlinearities becomes problematic as intersubband transitions couple only to the vertical electric field perpendicular to the QW plane. Furthermore, even at oblique incidence, the QW structure thickness is limited to 20-30 periods, which leads to short propagation distance and low efficiency of the nonlinear frequency conversion. One possible solution is to place a QW nanostructure in a high-Q nanocavity. This, however, requires certain fine tuning to ensure nonzero overlap of the three cavity modes involved in the SPDC process with the nonlinear region \cite{tokman2019-apl,tokman2022a}. Furthermore, high-Q cavities have low in- and outcoupling efficiency, and narrow bandwidth. 

A promising alternative approach is to integrate QW heterostructures with ultrathin metasurfaces that maximize the nonlinear overlap of the vertical electric fields with the nonlinear QW region and incoupling of the normally incident pump field without sacrificing the outcoupling efficiency of the signal and idler fields. This general approach has been demonstrated for the second and third harmonic generation processes in the mid-infrared spectral region \cite{lee2014,sarma2019,mekawy2021,park2024,wolf2015phased,sarma2022all} However, implementing it for the SPDC process and shifting the frequencies of the signal and idler fields into the near-infrared telecom range comes with a whole new set of challenges. The need to have the confining QW potential at least 2 eV deep combined with the requirement of bulk crystal transparency at the pump wavelength of 750 nm leaves the Nitride-based (e.g.GaN/AlN) QWs as the only practical material system. GaN/AlN QWs also offer a unique advantage: due to the presence of polarization charges and built-in electric field even a single well of symmetric AlN/GaN/AlN composition has a strong enough asymmetry to provide a high second-order intersubband nonlinearity, which is more than an order of magnitude higher than in a bulk crystal \cite{mundry2021nonlinear,hofstetter2007optically,nevou2006intersubband,kishino2002intersubband}. GaN/AlN-based heterostructures however have serious limitations for quantum applications as their growth is hindered by the inherent lattice-mismatch among the various nitrides. The strain induces defects and significantly limits the thickness of the films that can be grown. Furthermore, these defects result in interface roughening and non-uniformity of the films, which eventually lead to spectral broadening and reduction of the dipole moment  of the intersubband transitions.   
ScAlN/GaN/ScAlN heterostructures have recently emerged as a highly promising alternative  as by controlling the composition (Sc $\approx$ 0.14), the strain can be compensated and high quality heterostructures as thick as 500 nm can be grown \cite{Oana1,Oana2}.  These ScAlN/GaN/ScAlN heterostructures also maintain the high  conduction band offset (needed to scale the nonlinearity to telecom wavelengths) and retain the built-in electric field required to break the symmetry along the growth direction to  provide a high second-order intersubband nonlinearity. 

While having an ScAlN/GaN/ScAlN intersubband-based heterostructure satisfies the criteria of having a nonlinear medium with a giant nonlinearity, the net nonlinear signal from a metasurface is a result of a complicated light-matter interaction. The optimization of a metasurface for SPDC is a nontrivial task as it requires a balance of the nonlinear modal overlap, losses, and in/outcoupling efficiencies. One popular method of designing integrated nonlinear metasurfaces for such complicated light-matter interaction processes is inverse design via topology optimization, which has recently been applied in the mid-infrared \cite{mann2023, Stich_ACSnano_2025, molesky2018inverse}.  Here we utilize that technique and present the design, modeling, and detailed theory for an integrated nanophotonic system that includes a ScAlN/GaN/ScAlN multiple QW heterostructure with high intersubband nonlinearity integrated with a low-loss dielectric metasurface for highly efficient biphoton generation in the telecom range. As shown in Figure 1, we used topology optimization to inverse design a metasurface to maximize the nonlinear overlap of the pump, signal, and idler modes in the QW region.  Based on our theoretical model, the proposed  nanostructure of total thickness around 500 nm should generate a biphoton flux of $10^{10}$ s$^{-1}$ per unit cell of a metasurface in the near-infrared range around 1500 nm. Such biphoton generation rates are comparable with mode-matched high-Q nonlinear microresonators utilizing bulk III-V nonlinearity; e.g., \cite{zhao2022}. 

\section{Results}

\begin{figure}[h]
    \centering
    \includegraphics[width=0.75\columnwidth]{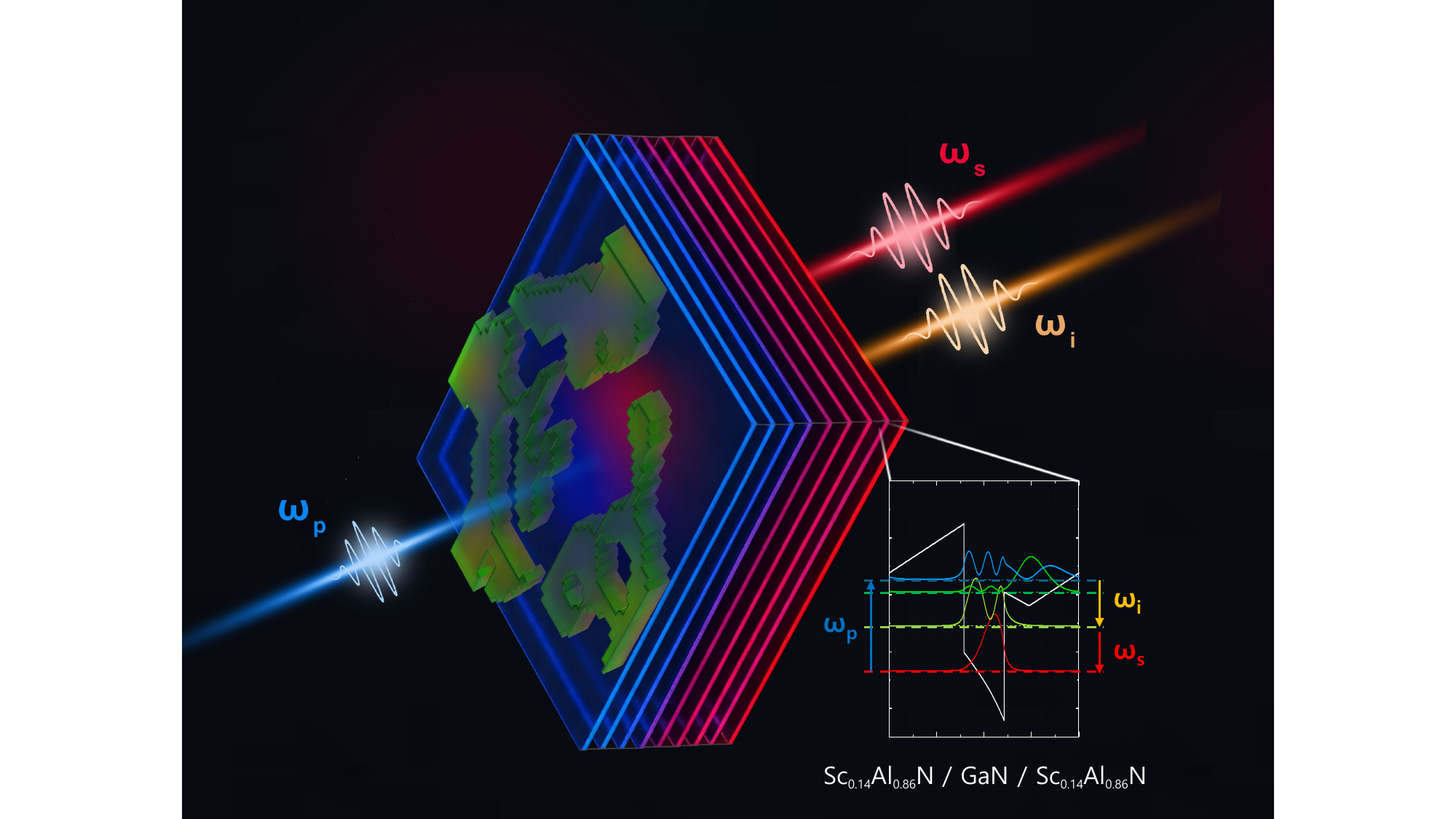}
    \caption{ Schematic of the process of entangled-photon generation from an inverse designed metasurface integrated with GaN/ScAlN quantum wells. A pump beam at frequency $\omega_p$ (pump, blue) excites a resonant mode of the metasurface, producing a strongly enhanced local field in the QW layer. The large second-order nonlinearity of the QWs then mediates spontaneous parametric down-conversion of the pump into a pair of lower-frequency photons at $\omega_s$ (signal, red) and $\omega_i$ (idler, orange). The inset shows the conduction-band profile of a single GaN/ScAlN QW period together with the probability density of the confined subband states used to realize the intersubband nonlinearity.
     }
	\label{fig1}
\end{figure}

\subsection{Nonlinear response of a dissipative electron system}

A general Heisenberg-Langevin framework for calculating the nonlinear optical response of an open dissipative electron system has been well developed; see, e.g., its application to parametric down-conversion in a cavity in our previous work \cite{graphene2013, tokman2017,tokman2022a}. 
 Consider electrons in conduction band subbands described by stationary states $|m\rangle$ and energy levels $W_m$. After introducing creation and annihilation operators of an electron, $\hat{a}^\dag_m|0\rangle = |m\rangle, \hat{a}_n|n\rangle = |0\rangle$, one can define the density operator as $\hat{\pi}_{mn} = \hat{a}^\dag_n\hat{a}_m$ which obeys the commutation relations $\lbrack \hat{\pi}_{qp}, \hat{\pi}_{mn}\rbrack = (\hat{\pi}_{mp}\delta_{nq} - \delta_{mp}\hat{\pi}_{qn})$. To describe the coupling of light to electrons or any quantum emitters distributed in space it is often convenient to introduce a coordinate-dependent density operator defined as
\begin{equation}
\label{1} 
\hat{\rho}_{mn}( \mb{r},t) = \frac{1}{\Delta V_r}\sum_{j}\hat{a}^\dag_{j;n}(t)\hat{a}_{j;m}(t)
\end{equation}
where the index j numerate individual electrons and the summation is carried over all electrons within a small volume $\Delta V_r$ in the vicinity of a point with radius-vector $ \mb{r}$. The density operator in Eq.~(\ref{1}) is normalized to the total electron density $N( \mb{r})$ according to $\langle \Psi_E | \sum_m \hat{\rho}_{mm} | \Psi_E \rangle = N( \mb{r})$ where $|\Psi_E\rangle $ is the initial wave function of the electron subsystem. The state vector $|\Psi_E\rangle $ should satisfy the indistinguishability of electrons which have quantum states overlapping in real space. Other than that, there is no need to know the specific form of  $|\Psi_E\rangle $, as long as we know the initial values of all relevant observables that are calculated by averaging with the Heisenberg density operator $\hat{\rho}_{mn}( \mb{r},t=0)$; for example, $\langle \Psi_E |  \hat{\rho}_{mm}( \mb{r},t=0) | \Psi_E \rangle = N_m( \mb{r},t=0)$ is the initial spatial density of populations (see, e.g., \cite{tokman2015}). 

Assuming that the operators in different points of space commute with each other, the commutation relations become
\begin{equation}
\label{2} 
\lbrack \hat{\rho}_{qp}( \mb{r}), \hat{\rho}_{mn}( \mb{r}^\prime)\rbrack = \delta( \mb{r}- \mb{r}^\prime)(\hat{\rho}_{mp}( \mb{r})\delta_{qn} - \delta_{mp}\hat{\rho}_{qn}( \mb{r})) .
\end{equation}
Using the density operator defined in Eq.~(\ref{1}), one can write the Heisenberg operator of any physical quantity $x( \mb{r},t)$ as
\begin{equation}
\label{3} 
\hat{x} = \sum_{n,m} x_{nm}\hat{\rho}_{mn}( \mb{r},t).
\nonumber
\end{equation}
In particular, the optical polarization which serves as a source in the field equations is given by $\hat{ \mb{P}}( \mb{r},t) = \sum_{n,m}  \mb{d}_{nm}\hat{\rho}_{mn}$ where $ \mb{d}_{mn}$ is the dipole matrix element.

Within the Heisenberg-Langevin approach \cite{SZ} and taking into account commutation relations (\ref{2}), the evolution of the density operator is described by the master equation,
\begin{equation}
\label{4} 
\dot{\hat{\rho}}_{mn} = -\frac{i}{\hbar}\left(\hat{h}_{mv}\hat{\rho}_{vn} - \hat{\rho}_{mv}\hat{h}_{vn}\right) + \hat{R}_{mn}(\hat{\rho}_{mn}) +
\hat{F}_{mn}.
\end{equation}
Here $\hat{h}_{nm} = W_n\delta_{nm} -  \mb{d}_{nm}\hat{ \mb{E}}( \mb{r},t)$ is the matrix element of the Hamiltonian operator $\hat{H} =
\hat{h}_{nm}\hat{a}^{\dag}_n\hat{a}_m$ describing interaction with the electric field $\hat{\bf{E}}( \mb{r},t)$ in the dipole approximation, $\hat{R}_{mn}$ the relaxation operator, for which we will choose the simplest form $\hat{R}_{m\ne n} = -\gamma_{mn}\rho_{mn}$,
$\hat{F}_{mn}$ the Langevin noise operator satisfying $\hat{F}_{mn} = \hat{F}^\dag_{nm}$ and $\langle \hat{F}_{mn} \rangle = 0$. The averaging
$\langle ... \rangle$ is taken both over the reservoir and over the initial state $|\Psi_E\rangle$ of the electron system.

To quantize the field, we need a proper normalization procedure. It is most convenient to treat the metasurface as a 3D cavity with well-defined modes. A typical quality factor in our design is $Q \sim 100$. We will take the total field $\hat{ \mb{E}}( \mb{r},t)$ as the sum of three cavity modes with frequencies related by the energy conservation in the parametric down-conversion process: 
\begin{equation}
\label{Eq:51}
	\omega_p=\omega_s+\omega_i.
    \nonumber 
\end{equation}
Here the pumping at frequency $\omega_p$ will be considered a classical coherent field, 
\begin{equation}
\label{Eq:52}
	\mb{E}_p=\mb{E}_p(\mb{r})e^{-i\omega_pt}+c.c.
\end{equation}
It is convenient to choose the phase of the pump mode so that its value  is real and positive. 
The field at signal and idler frequencies, $\omega_s$ and $\omega_i$ , will be the quantum field described by the operator 
\begin{equation}
\label{Eq:53}
	\mb{\hat{E}}=\sum_{\nu=s,i}\lbrack\mb{E}_\nu(\mb{r})\hat{c}_\nu+\mb{E}^*_\nu(\mb{r})\hat{c}^\dagger_\nu\rbrack,
\end{equation}
where $\hat{c}_\nu$ and $\hat{c_\nu}^\dagger$ are boson annihilation and creation operators. The functions $\mb{E}_{p,s,i}(\mb{r})$ in Eqs.~(\ref{Eq:52}) and (\ref{Eq:53}) are determined from solving the boundary-value problem of classical electrodynamics for a given geometry of the metasurface. 
The normalization of functions $\mb{E}_{\nu}(\mb{r})$ needs to be chosen in such a way that the commutation relation for boson operators $\hat{c}_\nu$ and $ \hat{c_\nu}^\dagger$ have a standard form $[\hat{c}_\nu,\hat{c_\nu}^\dagger]=\delta_{\nu\nu'}$. Following \cite{fain,tokman2015,tokman2016}, one can obtain 
\begin{equation} 
\label{3a} 
\int_V E_{\nu j}^*(\mb{r}) \frac{1}{2 \omega_{\nu}} \left[ \frac{ \partial \left( \omega^2 \varepsilon_{jk}(\omega,\mb{r}) \right) }{\partial \omega} \right]_{\omega = \omega_{\nu}}  E_{\nu k}(\mb{r})\, d^3r = 2 \pi \hbar \omega_{\nu}, 
\end{equation} 
where $\omega_{\nu}$ is the eigenfrequency of a cavity mode, $E_{\nu j}(\mb{r}) $ is a Cartesian component of the vector field $\mb{E}_{\nu}(\mb{r})$, $\varepsilon_{jk}(\omega,\mb{r}) $ is the dielectric tensor, and $V$ is a cavity volume (a quantization volume). 

Equation (\ref{3a}) is valid when the dissipation is weak enough. Specifically, the following three conditions have to be satisfied for a dissipation rate $\Gamma$ of a given cavity mode.  The first condition is obvious: $\Gamma \ll \omega$ has to be true for the frequencies of all modes involved in the parametric process. The second condition implies that the change of the Hermitian dielectric function $\varepsilon_{jk}(\omega)$ has to be small over the frequency interval of the order of $\Gamma$:  $|(\partial \varepsilon_{jk}(\omega)/\partial \omega) \Gamma | \ll   |\varepsilon_{jk}(\omega)|$.  The third condition states that the change in the derivative of $ \varepsilon_{jk}(\omega)$ which enters the expression for the EM energy density in Eq.~(\ref{3a}) must also be small:  $|(\partial^2 \varepsilon_{jk}(\omega)/\partial\omega^2) \Gamma |\ll  |(\partial \varepsilon_{jk}(\omega)/\partial \omega)|$.

In order to find commutation and correlation relations for the noise operators, we start from Eq.~(\ref{2}) which can be rewritten as
\begin{equation}
\label{5}
\lbrack \hat{\rho}_{mn}( \mb{r}), \hat{\rho}_{nm}( \mb{r}^\prime)\rbrack = \lbrack \hat{\rho}_{mn}( \mb{r}), \hat{\rho}^\dag_{mn}( \mb{r}^\prime)\rbrack = \delta( \mb{r}- \mb{r}^\prime)(\hat{\rho}_{nn}( \mb{r}) - \hat{\rho}_{mm}( \mb{r})) .
\end{equation}
From Eqs.~(\ref{4}) and (\ref{5}) one can obtain the commutation relation
\begin{equation}
\label{6}
\lbrack\hat{F}_{mn}( \mb{r},t), \hat{F}^\dag_{mn}( \mb{r}^\prime, t^\prime)\rbrack = 2\gamma_{mn}(\hat{\rho}_{nn} -
\hat{\rho}_{mm})\delta(t-t^\prime)\delta( \mb{r}- \mb{r}^\prime).
\end{equation}
Equation (\ref{6}) ensures that the commutator (\ref{5}) is preserved in the presence of relaxation $\gamma_{mn} \neq 0$.

We will also need a similar relation for the spectral components of the noise operator,
\begin{equation}
\label{7}
\hat{F}_{mn}(t, \mb{r}) = \int_{-\infty}^{\infty} \hat{F}_{\omega;mn}( \mb{r})e^{-i\omega t} d\omega, \hat{F}_{-\omega;mn} = \hat{F}^\dag_{\omega;nm}.
\end{equation}
Following \cite{graphene2013}, we arrive at 
\begin{equation}
\left.\begin{array}{cc}
\langle \hat{F}^\dag_{\omega;mn}( \mb{r}^\prime) \hat{F}_{\omega^\prime;mn}( \mb{r})\rangle & = \displaystyle  \frac{\gamma_{mn}}{\pi}\langle \hat{\rho}_{mm} \rangle \delta(\omega -
\omega^\prime)\delta( \mb{r} -  \mb{r}^\prime) \\
\langle \hat{F}_{\omega^\prime;mn}( \mb{r}) \hat{F}^\dag_{\omega;mn}( \mb{r}^\prime)\rangle &= \displaystyle   \frac{\gamma_{mn}}{\pi}\langle \hat{\rho}_{nn} \rangle \delta(\omega -
\omega^\prime)\delta( \mb{r} -  \mb{r}^\prime) \\
\end{array}\right\}
\label{fcorr2}
\end{equation}
If the system is in thermal equilibrium, i.e.$\langle\hat{\rho}_{nn}\rangle / \langle\hat{\rho}_{mm}\rangle = \exp{(\hbar\omega_{mn}/k_BT)}$, it follows from Eqs.~(\ref{fcorr2}) that the fluctuation component of the polarization $\hat{ \mb{P}}_L$ generated by the Langevin noise satisfies the fluctuation-dissipation theorem \cite{landau,rytov}:
$$
\frac{1}{2}\langle \hat{ \mb{P}}^\dag_{L;\omega^\prime}( \mb{r}^\prime)\hat{ \mb{P}}_{L;\omega}( \mb{r})+\hat{ \mb{P}}_{L;\omega}( \mb{r}) \hat{ \mb{P}}^\dag_{L;\omega^\prime}( \mb{r}^\prime)\rangle = \frac{\hbar}{\pi}{\rm Im}[\chi(\omega)]( n_T(\omega)+\frac{1}{2} )\delta(\omega-\omega^\prime)\delta( \mb{r}- \mb{r}^\prime)
$$
where
$$
\chi(\omega) = \frac{1}{\hbar}\sum_{m,n}\frac{| \mb{d}_{nm}|^2\langle\hat{\rho}_{nn}-\hat{\rho}_{mm}\rangle}{(\omega_{mn}-\omega)-i\gamma_{mn}}
$$
is the linear susceptibility of the medium, $n_T(\omega) = (e^{\hbar\omega/k_BT}-1)^{-1}$ is the average number of thermal quanta,
$$
\hat{ \mb{P}}_{L;\omega} = \sum_{m,n}\frac{ \mb{d}_{nm}\hat{F}_{\omega;mn}}{i(\omega_{mn}-\omega)+\gamma_{mn}}
$$
is the spectral component of the polarization $\hat{ \mb{P}}_L$.


\subsection{The second-order nonlinearity of GaN quantum wells with ScAlN barriers} 

The calculation of the nonlinear part of the polarization is more subtle because the Langevin noise terms at different frequencies are coupled together through the nonlinearity. Moreover, new noise terms are generated via the coherence excited by the pump field; see the rigorous analysis in \cite{tokman2017}. Since these are weak nonlinear effects, here we will ignore them assuming that the fluctuation component is related only to the linear absorption. Furthermore, since the strong pump field is classical, the resulting equations are linear with respect to quantized signal and idler, which greatly simplifies the analysis as compared to the quantized pump case; see, e.g., \cite{tokman2022a, tokman2022b, tokman2023a}. Finally, we will neglect band nonparabolicity assuming that it gives rise to a smaller spectral broadening as compared to the homogeneous broadening due to energy and phase relaxation.  The nonparabolicity is straightforward to take into account if needed by performing proper summation over electron momentum states when calculating the polarization. After these approximations the QW system becomes effectively a three-level medium, similar to the one considered, e.g., in \cite{nefedkin2021}. The strong pump field can be taken into account exactly in all orders, as was done, e.g., in 
\cite{nefedkin2021}. Note that for these ScAlN/GaN-based heterostructures an inversion symmetry breaking effect induced by a strong built-in electric field enables very high second-order nonlinearity even in a single QW. Figure 2(a) shows k.p band structure
calculations of a single strain compensated (Sc composition $\approx$ 0.14 \cite{Oana1, Oana2}) GaN quantum well. Due to polarization-induced band tilting that breaks inversion symmetry, as shown later, this simple QW design can exhibit $\chi^{(2)}$ $\approx$ 10 nm/V at 1.55 $\mu$m. Coupled double-well designs provide additional tunability of well depth, width, and inter-well coupling, increasing dipole matrix elements and boosting $\chi^{(2)}$ to $\approx$ 20 nm/V (Fig. 2(b)).  We want to point out that unlike previous approach of doping the wells, in these QW designs, we have further enhanced symmetry breaking through single-side n-type modulation doping of the barriers (sheet density $ 2.4\times 10^{13}$ cm$^{-2}$), which transfers carriers into the wells and induces asymmetric band bending to increase $\chi^{(2)}$.


\begin{figure}[h]
    \centering
    \includegraphics[width=0.9\columnwidth]{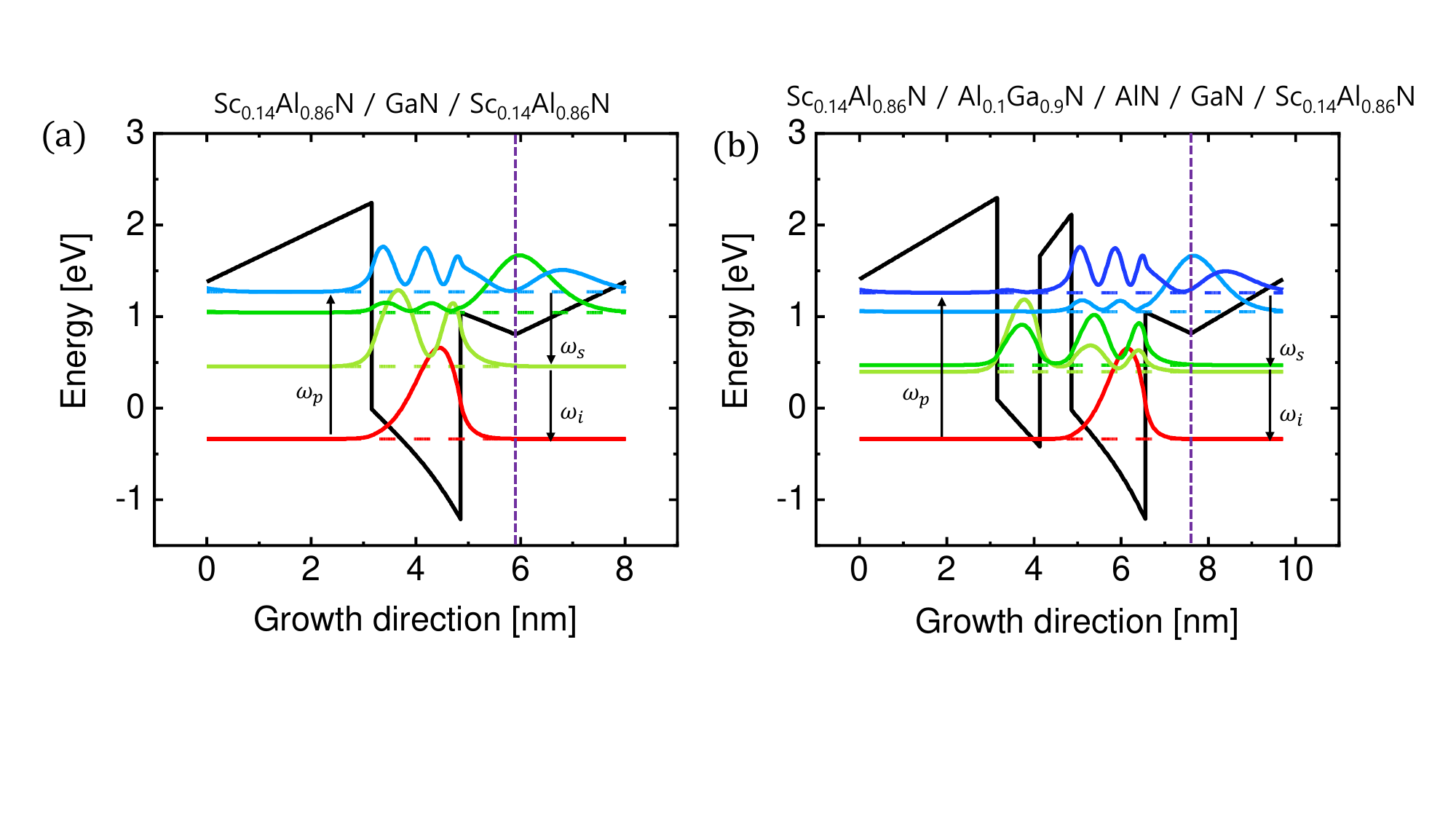}
    \caption{8-band k.p band structure calculations of conduction band edge profiles of a single heterostructure period and the confined energy levels involved in SPDC in (a) a single QW consisting of 7 monolayers of GaN QW with 6 nm Sc$_{0.14}$Al$_{0.86}$N barriers, and (b) a double QW composed of 4 monolayers of GaN, 3 monolayers of AlN, and 7 monolayers of GaN, also confined by 6 nm Sc$_{0.14}$Al$_{0.86}$N  barriers. An N-type delta-doping layer with a sheet density $ 2.4\times 10^{13}$ cm$^{-2}$ is introduced 1 nm away from the GaN quantum well, as indicated by the vertical purple dashed line.}
	\label{fig2}
\end{figure}

Below we will give approximate analytic expressions for the nonlinear response when the pump is not strong enough to cause significant optical pumping and redistribution of populations across excited subbands. This would correspond to the second-order nonlinearity, i.e., the component of the polarization which scales as 
$ \mb{P}^{(2)}  = \chi^{(2)}(\omega_s = \omega_p - \omega_i) \mb{E_p} \mb{E_i^*} + \chi^{(2)}(\omega_i = \omega_p - \omega_s) \mb{E_p} \mb{E_s^*} + c.c.$; see the exact definition in Eq.~(\ref{eqn4}) below.  
 A strong optical pumping exceeding the saturation intensity would be not an optimal regime as it not only reduces the SPDC efficiency but also adds noise due to spontaneously emitted uncorrelated photons. In the case of moderate pumping intensity the subband populations remain in equilibrium defined by doping and temperature. As an added bonus, we do not need to take into account the dynamic many-body Coulomb effects caused by redistribution of the populations. The space charge induced energy shift at high doping densities can be significant, but it has been already included in the equilibrium bandstructure simulations. 

The resulting second-order nonlinear susceptibility obtained from solving Eq.~(\ref{4}) is particularly simple in the two limiting cases. The first case is when the signal and idler frequencies are sufficiently different and we assume the rotating-wave approximation in which each field is coupled only to one transition closest to resonance. In this case the second-order susceptibility at signal and idler frequencies is given by 
\begin{equation}
\label{chi2s} 
\chi^{(2)}(\omega_s = \omega_p - \omega_i) = \frac{d_{12} d_{23} d_{13}}{\hbar^2 \left(\gamma_{32} + i  (\omega_{32} - \omega_s)\right)} \left[ \frac{n_1 - n_3}{ \gamma_{31} + i  (\omega_{31} - \omega_p)} + \frac{n_1 - n_2}{ \gamma_{21} - i  (\omega_{21} - \omega_i)} \right]
\end{equation}
and 
\begin{equation}
\label{chi2i} 
\chi^{(2)}(\omega_i = \omega_p - \omega_s) = - \frac{d_{12} d_{23} d_{13}}{\hbar^2 \left(\gamma_{21} + i  (\omega_{21} - \omega_i)\right)} \left[ \frac{n_1 - n_3}{ \gamma_{31} + i  (\omega_{31} - \omega_p)} + \frac{n_2 - n_3}{ \gamma_{32} - i  (\omega_{32} - \omega_s)} \right], 
\end{equation}
where $d_{mn}$, $m,n = 1,2,3$ are dipole matrix elements for the optical transitions between levels $m$ and $n$; $\gamma_{mn}$ is the decoherence rate (half-width at half-maximum) of the optical transition $m \rightarrow n$ at frequency $\omega_{mn} = (W_m - W_n)/\hbar$; $n_{1,2,3}$ are electron densities in cm$^{-3}$ at subbands $1,2,3$.  Since only the z-polarized fields are coupled to intersubband transitions, where $z$ is the QW growth direction, only $z$ components of the dipole moments are nonzero and the expressions above are actually $\chi^{(2)}_{zzz}$ components of the tensor.

For large enough detunings one can neglect $\gamma_{mn}$, and expressions (\ref{chi2s}) and (\ref{chi2i}) become real and equal to each other as they should be from the symmetry properties which ensure that Manley-Rowe relationships are satisfied in a conservative system \cite{keldysh, te2011}. 

When the signal and idler frequencies are close to each other, one should take into account the coupling of each field to both optical transitions, $1 \rightarrow 2$ and $2 \rightarrow 3$. The resulting expressions become a bit cumbersome, but a simplification occurs in the case of degenerate SPDC when the signal and idler frequencies are equal to each other. In this case, as one could guess, the resulting $\chi^{(2)}$ is the sum of expressions (\ref{chi2s}) and (\ref{chi2i}): 
\begin{equation}  
\chi_s^{(2)}  =  \displaystyle  \frac{d_{12} d_{23} d_{13}}{\hbar^2} \left( \frac{1}{\left(\gamma_{32} + i  (\omega_{32} - \omega_s)\right)} \left[ \frac{n_1 - n_3}{ \gamma_{31} + i  (\omega_{31} - \omega_p)} + \frac{n_1 - n_2}{ \gamma_{21} - i  (\omega_{21} - \omega_s)} \right] 
\right. 
\nonumber 
\end{equation}
\begin{equation} 
\left. 
\displaystyle    -  \frac{1}{ \left(\gamma_{21} + i  (\omega_{21} - \omega_s)\right)} \left[ \frac{n_1 - n_3}{ \gamma_{31} + i  (\omega_{31} - \omega_p)} + \frac{n_2 - n_3}{ \gamma_{32} - i  (\omega_{32} - \omega_s)} \right]  \right). 
\label{chi2tot} 
\end{equation} 
Note the negative interference between the two SPDC channels described by the first and second line in Eq.~(\ref{chi2tot}) which results in their partial cancellation. 

Figure 3 shows two examples of the $|\chi_s^{(2)}|$ spectra calculated numerically for the QWs in Fig. 2(a) and (b),  without making the above approximations, and assuming the linewidths $\gamma_{21} = \gamma_{32} = \gamma_{31}= 46.5$ meV, based on recent experimental reports \cite{Oana1, Oana2}. The electron wavefunctions, subband energies, and transition matrix elements were calculated by performing 8-band k.p band structure calculations with the NextNano software. The intersubband dipole matrix elements  between subbands were extracted by averaging over all dipole-allowed transitions between the degenerate states (see Supporting Information) and the characteristic magnitude of $|\chi_s^{(2)}|$ is nearly 10 nm/V at the peak in Fig.~3(a) and still around 5 nm at the detuning of $2\times \gamma_{32}$ from the ISB resonance. This is lower than mid-infrared resonant ISB nonlinearities for GaAs- or InP-based asymmetric coupled QWs  but still two orders of magnitude higher than the bulk nonlinearity of the GaN crystal and in fact any bulk nonlinear crystal utilized for SPDC in the telecom range. 

\begin{figure}[h]
    \centering
    \includegraphics[width=0.9\columnwidth]{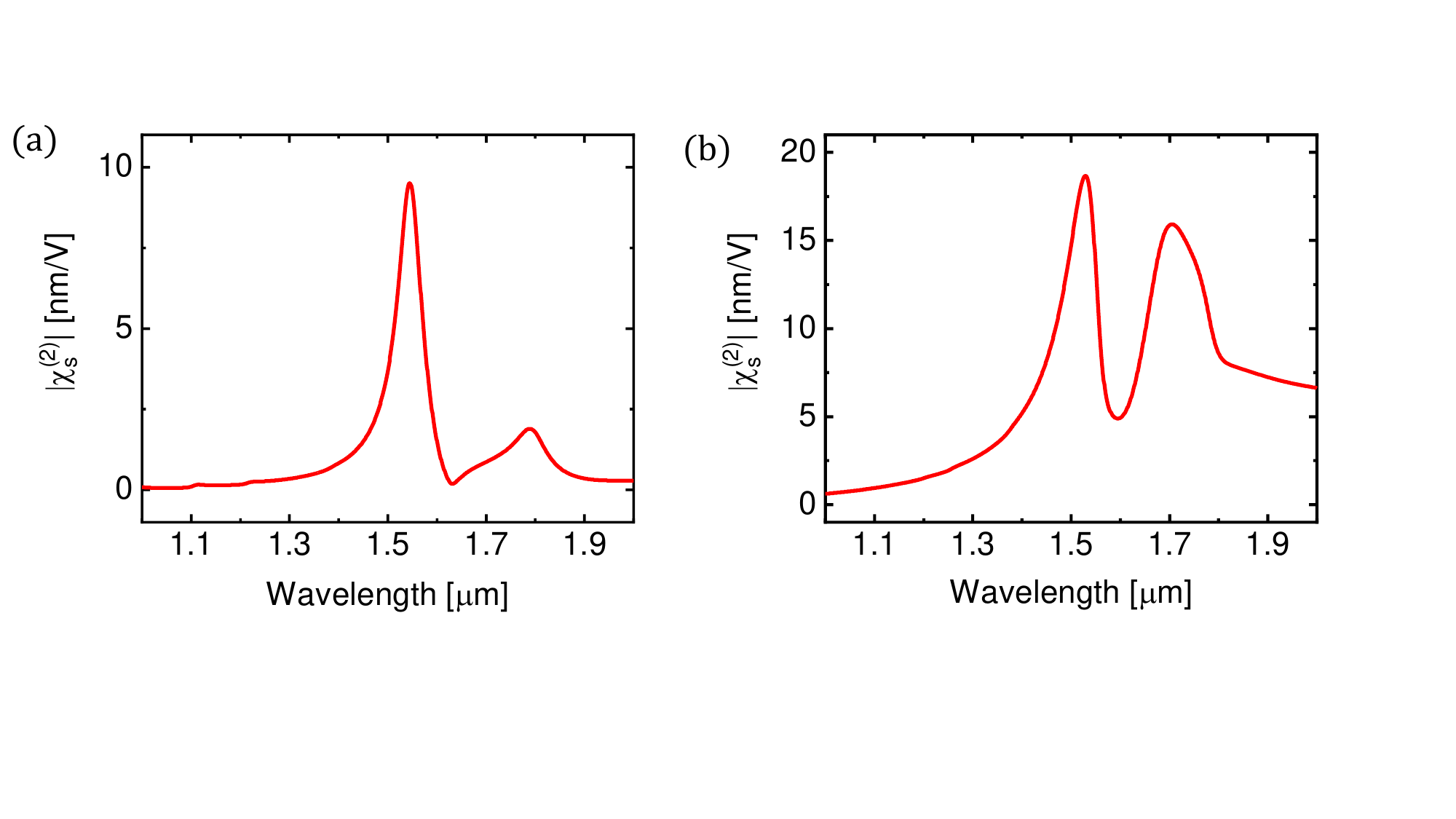}
    \caption{The spectrum of $|\chi_s^{(2)}|$ as a function of the signal field wavelength. Panel (a): for the structure in Fig.~2(a); Panel (b): for the structure in Fig.~2(b). In both (a) and (b) an N-type delta-doping layer with a sheet density $ 2.4\times 10^{13}$ cm$^{-2}$ is introduced 1 nm away from the GaN quantum well.}
	\label{fig3}
\end{figure}


\subsection{Parametric down-conversion efficiency and signal flux }

The second-order nonlinearity of the QWs gives rise to the nonlinear polarization at signal and idler frequencies. The excitation equations for the cavity modes derived from the operator-valued Maxwell's equations \cite{fain} take the form 
\begin{equation}
\label{Eq:54}
    \dot{\hat{c}}_\nu+i\omega_\nu\hat{c}_\nu=
-\frac{i}{\hbar\omega^2_\nu}\int\limits_V\ddot{\hat{\mb{P}}}_{ex}(\mb{r},t)\mb{E}^*_\nu(\mb{r})d^3r,
\end{equation}
where the external polarization is the sum of the components due to the nonlinearity, linear (proportional to the field) absorption in the material, and fluctuations (Langevin noise term):
\begin{equation}
\label{pol}
\hat{\mb{P}}_{ex} = \hat{\mb{P}}_{NL} + \hat{\mb{P}}_{lin} + \hat{\mb{P}}_{L}.
\end{equation}
As a reminder, we follow the standard approach in which the fluctuation term is related with only the linear dissipation term. Of course the nonlinearity parametrically couples the Langevin terms at different frequencies. However, this can be considered a small contribution.   

The positive-frequency $\propto e^{-i \omega t}$ parts of the corresponding components of the polarization can be expressed through modal losses and fluctuations as
\begin{equation}
\label{lin}
-\frac{i}{\hbar\omega^2_\nu}\int\limits_V\ddot{\hat{\mb{P}}}_{\nu, lin}^{(+)}(\mb{r},t)\mb{E}^*_\nu(\mb{r})d^3r = - \Gamma_{\nu} \hat{c}_\nu,
\end{equation}
\begin{equation}
\label{noise}
-\frac{i}{\hbar\omega^2_\nu}\int\limits_V\ddot{\hat{\mb{P}}}_{\nu, L}^{(+)}(\mb{r},t)\mb{E}^*_\nu(\mb{r})d^3r = - \hat{L}_{\nu} e^{-i\omega_\nu t},
\end{equation}
where $\hat{c}_\nu \propto e^{-i\omega_\nu t}$ and modal losses $ \Gamma_{\nu}$ can be calculated for a given metasurface geometry and field distribution. 

The positive-frequency part of the nonlinear polarization is 
\begin{equation}
\hat{\mathbf{P}}_{NL}^{(\mathbf{+})} = \hat{\chi}^{(2)}_s(\mathbf{r}, \omega_{s} = \omega_{p} - \omega_{i} )\mathbf{E}_{p}(\mathbf{r})\mathbf{E}_{i}^{\mathbf{*}}(\mathbf{r})e^{-i\omega_pt} \hat{c}_{i}^{\dagger} +  \hat{\chi}^{(2)}_i(\mathbf{r}, \omega_{i} = \omega_{p} - \omega_{s} )\mathbf{E}_{p}(\mathbf{r})\mathbf{E}_{s}^{\mathbf{*}}(\mathbf{r})e^{-i\omega_pt} \hat{c}_{s}^{\dagger} ,
\label{eqn4}
\end{equation}
where $\hat{c}_{s,i}^{\dagger} \propto e^{i\omega_{s,i} t}$.

When the losses and corresponding fluctuations are neglected,  Eqs.~(\ref{Eq:54}) and (\ref{eqn4}) give the following parametrically coupled equations in the rotating wave approximation:  
\begin{equation}
\label{Eq:59}
     \dot{\hat{c}}_s+i\omega_s\hat{c}_s+ g_s e^{-i\omega_pt}\hat{c}_i^\dagger=0,\hspace{5 mm}
          \dot{\hat{c}}_i^\dagger-i\omega_i\hat{c}^\dagger_i+ g_i^* e^{+i\omega_pt}\hat{c}_s=0
\end{equation}
where we introduce the nonlinear overlap integrals of the cavity modes with the second-order nonlinear susceptibility  distribution, integrated over the cavity volume: 
\begin{equation}
g_s = -\frac{i}{\hbar}  \int_{V}{\mathbf{E}_{s}^{\mathbf{*}}(\mathbf{r})\hat{\chi}_s^{(2)}(\mathbf{r})\mathbf{E}_{p}(\mathbf{r})\mathbf{E}_{i}^{\mathbf{*}}(\mathbf{r})d^{3}r},
\label{eqn7}
\end{equation}
\begin{equation}
g_i = -\frac{i}{\hbar}  \int_{V}{\mathbf{E}_{i}^{\mathbf{*}}(\mathbf{r})\hat{\chi}_i^{(2)}(\mathbf{r}) \mathbf{E}_{p}(\mathbf{r})\mathbf{E}_{s}^{\mathbf{*}}(\mathbf{r})d^{3}r}.
\label{eqn8}
\end{equation}
Since only the z-polarized fields are coupled to intersubband transitions, where $z$ is the QW growth direction, we can take $z$ components of all fields in these integrals and the $\chi^{(2)}_{zzz}$ component of the tensor.

Now we take into account linear absorption and radiative losses within the Heisenberg-Langevin formalism. 
 Introducing  operators of slowly varying field amplitudes, namely $\hat{c}_{s,i}=\hat{c}_{0s,i}(t)e^{-i\omega_{s,i}t}$, $\hat{c}_{s,i}^{\dagger}=\hat{c}^\dagger_{0s,i}(t)e^{+i\omega_{s,i}t}$, we obtain from Eqs.~(\ref{Eq:54})-(\ref{Eq:59}) the following equations:
	\begin{equation}
	\label{Eq:66}
\left.
\begin{array}{r}
\dot{\hat{c}}_{0s}+\Gamma_s\hat{c}_{0s}+ g_s\hat{c}_{0i}^\dagger=\hat{L}_s \\
\dot{\hat{c}}_{0i}^\dagger+\Gamma_i\hat{c}_{0i}^\dagger + g_i^* \hat{c}_{0s}=\hat{L}_i^\dagger
\end{array} \right\},
\end{equation}
where $\hat{L}_{s,i}$ are the Langevin noise operators,  $\Gamma_{s,i}=\Gamma_{r(s,i)}+\Gamma_{\sigma(s,i)}$, and the coefficients $\Gamma_{r(s,i)}$ and $\Gamma_{\sigma(s,i)}$ denote, respectively, radiative losses due to the outcoupling of radiation from the cavity and absorption losses due to intracavity absorption. They need to be calculated separately for a given metasurface geometry, material absorption, and spatial distribution of the signal and idler modes.

 We show in \cite{tokman2019-apl}  that to preserve commutation relations $[\hat{c}_{0i},\hat{c}_{0i}^\dagger]=[\hat{c}_{0s},\hat{c}_{0s}^\dagger]=1$ at $\Gamma_{s,i}\neq0$ the noise operators in the right-hand side of Eqs.~(\ref{Eq:66}) should satisfy the same commutation relations as in the case of one quantum oscillator \cite{SZ,gardiner,tokman2018} and they should also commute with each other: 
\begin{equation}
\label{Eq:67} 
	\left[ \hat{L}_s(t'),\hat{L}_s^\dagger(t)\right]=2\Gamma_s\delta(t-t'),\hspace{5 mm}
\left[ \hat{L}_i(t'),\hat{L}_i^\dagger(t)\right]=2\Gamma_i\delta(t-t'),\hspace{5 mm}
	\left[ \hat{L}_s(t'),\hat{L}_i^\dagger(t)\right]=0.
\end{equation}
The fact that commutation relations are the same for one quantum oscillator and for two (or more) interacting oscillators is expected, since the processes within the Hamiltonian dynamics do not affect the commutators. Noise correlators can be defined by generalizing the simplest expression in \cite{SZ} to the case of two absorbers with different temperatures: 
	\begin{equation}
	\label{Eq:68}
\left.
\begin{array}{r}
\langle\hat{L}^\dagger_s(t)\hat{L}_s(t')\rangle=2\left[\Gamma_{\sigma s}n_{T_\sigma}(\omega_s)+\Gamma_{r s}n_{T_r}(\omega_s)\right]\delta(t-t') \\
\langle\hat{L}^\dagger_i(t)\hat{L}_i(t')\rangle=2\left[\Gamma_{\sigma i}n_{T_\sigma}(\omega_i)+\Gamma_{r i}n_{T_r}(\omega_i)\right]\delta(t-t')
\end{array} \right\}
\end{equation}
where $n_{T_{r,\sigma}}$ is the average number of thermal photons  at temperature $T_{r,\sigma}$;  $T_r$ and $T_{\sigma}$ denote the temperature outside and inside the cavity, respectively. Expressions (\ref{Eq:68}) imply that dissipative and radiative noises are not correlated. If the dissipation is only due to resonant absorption in QWs, all the expressions for modal losses and Langevin sources in the field equations can be obtained from calculating the polarization (\ref{pol}) with the help of the master equation (\ref{4}); see \cite{graphene2013,tokman2017}.  

The general solution of Eq.~(\ref{Eq:66}) which includes both stimulated and spontaneous terms is rather cumbersome and is presented in the Supporting Information. Here we only give the result for the regime of SPDC when the parametric growth rate is much lower than losses, $ g < \Gamma_s$. We will also consider degenerate SPDC for simplicity and 
assume that  $g = \sqrt{g_s g_i^*} = |g_s| =  |g_i|$ is real and positive. In the stationary limit, when $(\Gamma_s -  g) t \rightarrow \infty$, the radiated signal power  $P_{rs} \approx 2 \Gamma_{rs} \hbar \omega_s n_s$  becomes 
\begin{equation}
\label{spont}
P_{rs} = \hbar \omega_s \frac{2 \Gamma_{rs} \Gamma_s}{\Gamma_s^2 -  g^2} \left[\Gamma_{\sigma s} n_{T_{\sigma}} (\omega_s) + \Gamma_{rs} n_{T_r} (\omega_s) \right] + \hbar \omega_s \frac{\Gamma_{rs}  g^2}{\Gamma_s^2 -  g^2}.
\end{equation} 
The first term on the right-hand side of Eq.~(\ref{spont}) is due to the thermal emission modified by the parametric decay of the pump photons. The second term originates from the parametric decay of the pump photons under the action of vacuum fluctuations of the intracavity field; this is a purely quantum effect.  Thermal effects can be neglected if 
$$ \frac{\Gamma_s \Gamma_{(\sigma,r)s}}{ g^2} \frac{2}{\exp\left(\hbar\omega_s/(k_B T_{\sigma,r}) \right) -1 } \ll 1. 
$$ 

The results in Supporting Information and the SPDC limit given by Eq.~(\ref{spont}) provide the dependence of the parametric signal from all relevant parameters including the nonlinear overlap of the pump, signal, and idler modes, outcoupling of the signal and idler photons, dissipation, and fluctuation effects.   Equation (\ref{spont}) determines the spontaneous parametric signal emitted from a metasurface against the background of noise created by both thermal radiation from a metasurface and reemission of thermal photons coupled into a metasurface from the outside. The background noise depends on the sample temperature and the environment temperature.  

\subsection{Inverse design of dielectric metasurface for enhancing SPDC generation rates}

To maximize the biphoton generation rates, we designed a dielectric metasurface that simultaneously supports tightly confined cavity modes at both the pump ($\omega_p$) and the signal ($\omega_s$) frequencies. From a photonic design perspective, it is desirable to realize cavity modes that exhibit strong spatial confinement (small effective mode volume $V$) and large nonlinear modal overlap $g$, rather than radiatively limited modes characterized by high radiation loss rates ($\Gamma_{rs}$). The former not only enhances the nonlinear conversion efficiency through increased field intensity and overlap but also provides improved tolerance against fabrication imperfections. Although an exact optimization of the signal power would, in principle, require a full solution of the nonlinear Maxwell’s equations, as discussed in \cite{Stich_ACSnano_2025}, an equivalent and computationally efficient formulation can be derived by solving a set of coupled linear scattering problems at  frequencies $\omega_p$ and $\omega_s$. For simplicity, we restrict our analysis to the degenerate case where the signal and idler frequencies are identical, $\omega_s = \omega_i = \omega$, and the pump frequency satisfies $\omega_p = 2\omega$.

To optimize the cavity mode spatial overlap, which we parameterize in this section as 
\begin{equation}
\beta = \frac{1}{4}\frac{\int \varepsilon_0\sum_{ijk}\chi^{(2)}_{ijk} \left( {E}_{s,i}^{*}{E}_{s,j}^{*}{E}_{p,k}+ {E}_{s,i}^{*}{E}_{p,j}{E}_{s,k}^{*}\right) \ d\vb{r} }{\left( \int \varepsilon_0\varepsilon_1|\mathbf{E}_s|^2 \ d\vb{r}\right) \sqrt{\int \varepsilon_0\varepsilon_2|\mathbf{E}_p|^2\ d\vb{r}} },
\end{equation} 
we begin by setting up a dipole source at signal frequency $\omega_s$ at the center of the nonlinear material. We note that $\beta$ is proportional to the nonlinear modal overlap $g$ as described above. Now, instead of optimizing for Purcell Enhancement for two otherwise uncorrelated individual resonances at $\omega_s$ and $\omega_p$, we consider not a dipole source but an extended source $\vb{J}_{ind} \sim \vb{E}_s^2$ at pump frequency $\omega_p$, and optimize for a single combined objective $\Phi_{ps} = - \re{\int_\Omega \vb{J}_{ind}^* \cdot \vb{E}_p \ d\vb{r}}$ which incorporates $\beta$ by coupling two scattering problems at $\omega_s$ and $\omega_p$ \cite{Mohajan_OE_2023,Lin_Optica_2016,Stich_ACSnano_2025}. Hence, $\Phi_{ps}$ yields precisely the $\beta$ parameter along with any resonant enhancement factors $\sim Q/V$ in $\vb{E}_s$ and $\vb{E}_p$. Intuitively, $\vb{J}_{ind}$ can be thought of as a nonlinear polarization current induced by $\vb{E}_s$ in the presence of the second-order susceptibility tensor $\chi^{(2)}$. Therefore, the scattering problem with the objective of maximizing the spatial mode overlap between the two frequencies take the form:
\begin{equation}\label{eq:opt1}
\begin{aligned}
\max_{\overline{\rho}} \quad &  \Phi_{ps} =  -\frac{1}{2} \re{\int_\Omega \vb{J}_{ind}^* \cdot \vb{E}_p \ d\vb{r}} \\
\textrm{s.t.} \quad & \mathbb{M}_{s} \vb{E}_s = \overline{\chi}_{1}\vb{E}_s +  \frac{i}{\omega_s \varepsilon_0} \vb{J}_s + \frac{i}{\omega_s\varepsilon_0} \vb{J}_{dipole}\\
&\vb{J}_{ind} = -i\omega_p\varepsilon_0\overline{\chi}^{(2)}\vb{E}_s \vb{E}_s\\
&\mathbb{M}_{p} \vb{E}_p = \overline{\chi}_{2} \vb{E}_p + \frac{i}{\omega_p \varepsilon_0}\vb{J}_{ind}\\
&\overline{\chi} = \chi \overline{\rho}, \qquad \overline{\rho} \in [0,1].  
\end{aligned}
\end{equation}
Here we use the SI units where $\varepsilon_0$ and $\mu_0$ are the vacuum permittivity and permeability respectively, the superscript $( ^*)$ denotes conjugated vector fields, and $\overline{\chi}^{(2)}\vb{E}_s \vb{E}_s$ is the action of the rank-$3$ tensor field $\overline{\chi}^{(2)}(\vb{r})$ on different polarizations of vector field $\vb{E}_s(\vb{r})$. The $e^{-i\omega t}$ time convention is used to represent harmonic fields, with the vacuum Maxwell operator given by $\mathbb{M}_{l} = \left(-\mathbb{I}+ \frac{1}{\omega_l^2 \varepsilon_0}\curl\frac{1}{\mu_0}\curl\right), \ l={s,p}$, where $\mathbb{I}$ is the identity operator. We note that $\vb{J}_{ind}$ is the induced current resulting from nonlinear up-conversion which can be treated as a free source under the undepleted pump approximation.
Furthermore, to account for the trade offs between a high quality factor Q of a mode and higher spatial modal overlap, the quality factors of the modes are constrained to a smaller value promoting higher overlap. As shown later, maximizing overlap is crucial as it is the dominant contributor to the biphoton generation rate.

To maximize the coupling to plane waves for incident pump light and the outcoupling of the generated  signal, we introduce auxiliary objectives  $\Phi_{p,s}$ which help converting polarization of the incident field, and concentrate light in the quantum well region:
\begin{equation}\label{eq:opt2}
\begin{aligned}
\max_{\overline{\rho}} \quad &  \Phi_{p,s} =  \int_{QW} \abs{E_{z_{p,s}}}^2 \ d\vb{r} \\
\textrm{s.t.} \quad & \mathbb{M}_{p,s} \vb{E}_{p,s} = \overline{\chi}_{p,s}\vb{E}_{p,s} +  \frac{i}{\omega_s \varepsilon_0} \vb{J}_{inc_{p,s}} \\
&\overline{\chi} = \chi \overline{\rho}, \qquad \overline{\rho} \in [0,1].  
\end{aligned}
\end{equation}
Lastly, we define an objective that takes into account both the coupling and the cavity overlap factor, namely $\Phi_p\Phi_{ps}\Phi_s$, and optimize for the structural degrees of freedom as one single optimization ( see Supporting Information for details).

\begin{figure}[hbtp]
    \centering
    \includegraphics[width=0.6\columnwidth]{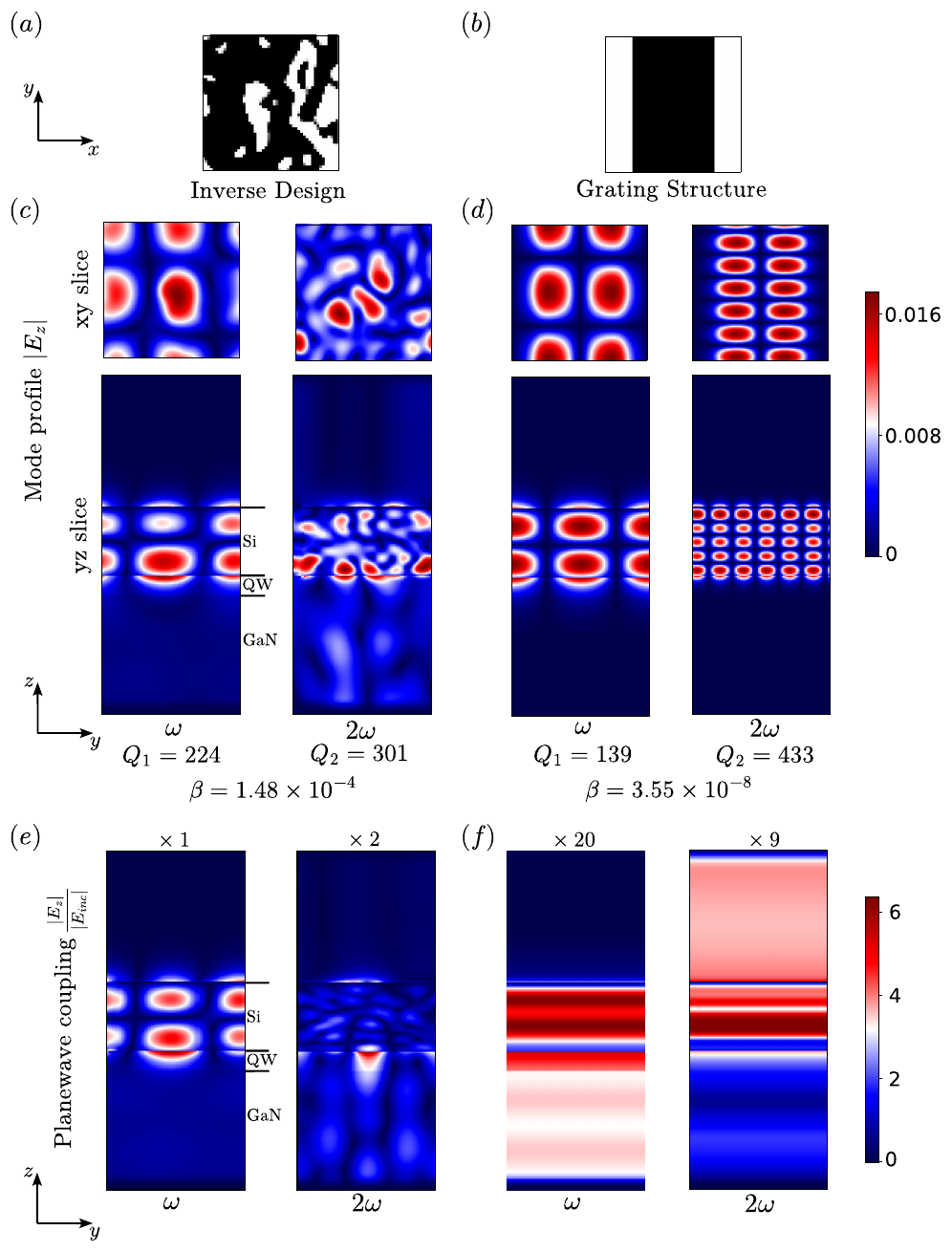}
    \caption{(a, b)  In-plane (x-y) spatial profile of one period of the inverse-designed metasurface (a) and a dielectric grating (b). The black region corresponds to presence of material (silicon) and white region corresponds to free space. The metasurface and gratings are assumed to be an infinite periodic array of the structures shown in (a) and (b). (c, d) Numerically calculated absolute value of the z-component of the electric field in the central x-y plane of the nonlinear multi-QW region and y-z plane at the center of the unit cell for the modes at the pump (2$\omega$) and signal frequencies ($\omega$) for the inverse designed metasurface (c) and dielectric grating (d). The corresponding quality factors (Q) of the modes for both the structures are mentioned in the figures. (e,f) Numerically calculated absolute value of the z-component of the electric field normalized with respect to the incident field $E_{inc}$ in the  y-z plane at the center of the unit cell for planewave excitation at (2$\omega$) and signal frequencies ($\omega$) for the inverse-designed metasurface (a) and dielectric grating (b). The inverse-designed structure couples significantly better to the modes shown in (c) compared to the dielectric gratings (d).}
	\label{fig4}
\end{figure}

Fig.~(4) shows a schematic of a cell of the optimized periodic structure obtained from inverse design comprising an amorphous silicon metasurface on top of the nonlinear QW layer which sits on GaN buffer layer. The buffer layer is included to mimic realistic growth conditions. For inverse design, we assumed that only the silicon layer can be inverse-designed for the optimization and  the QW layer is left untouched. For comparison, we used a periodic grating structure that was optimized for the highest spatial modal overlap. Fig.~4 shows the modal field profile of $\abs{E_z}$ at both the signal and pump frequency for both the grating and the inverse designed metasurface. The normalized cavity mode spatial overlap integral factor $\beta$ is $\approx 10^{-4} \frac{\chi^{(2)}}{\sqrt{\epsilon_0 \lambda^3}}$ for the inverse-designed structure compared to $\approx 10^{-8} \frac{\chi^{(2)}}{\sqrt{\epsilon_0 \lambda^3}}$ for the gratings, which is a four orders of magnitude improvement for the overlap integral.

\subsection{Generation rate of biphotons}

Finally, we estimate the biphoton generation rate for the inverse designed metasurface introduced in Fig.~\ref{fig4}.

For the optimal metasurface described in the previous section, the signal losses are dominated by the radiative outcoupling losses, $\Gamma_{s} \approx \Gamma_{rs}$, the quality factor of the mode of the metasurface at the signal frequency is $ Q \approx 100$, and the nonlinear overlap factor $g \ll \Gamma_s$. Therefore, the generation rate of biphotons is given by 
\begin{equation}
\label{rate}
R = \frac{P_{rs}}{\hbar \omega_s} \approx \frac{g^2}{\Gamma_{rs}} \approx \frac{g^2}{\omega_s} Q.
\end{equation}

As a numerical example, we consider one unit cell of the metasurface, with an area of $A = 885\times 885$ nm$^2$ and thickness of 555 nm, which includes the dielectric layer and the QW layer. Using the nonlinear overlap integral calculated for the metasurface in Fig.~4 with the value of $\chi^{(2)} \approx 10$ nm/V gives the biphoton generation rate $R \sim 10^{10}$ s$^{-1}$ per 1 Watt of incident pump power from one cell of the metasurface. For reference, 1 Watt of incident power would correspond to the z-component of the pump field $E_p \sim 10^7$ V/m in the unit cell volume, as follows from the metasurface simulations.  Also, 1  Watt of incident power per unit cell corresponds to about $10^8$ W/cm$^2$ of intensity. This is still below the onset of saturation and the optical damage which is in the GW/cm$^2$ range for this QW system. The scaling of the biphoton generation rate with the pump power is obviously linear below saturation.  Also, the above value of the biphoton generation rate was calculated per one unit cell and will increase linearly with the number of illuminated unit cells.


\section{Discussion}

 
 We have presented a microscopic quantum–optical theory and a concrete nanophotonic design for generating entangled photons in the telecommunications band using GaN/ScAlN multiple quantum wells integrated with an inverse designed dielectric metasurface.  
By combining a rigorous Heisenberg–Langevin formalism, including dissipation, fluctuations, and parametric amplification of both thermal and vacuum noise, with full-wave electromagnetic simulations and inverse design, we show that intersubband nonlinearities in ScAlN-based heterostructures can be harnessed efficiently under normal-incidence pumping, despite their intrinsically $z$–polarized selection rules.  
A key outcome of this analysis is the identification of GaN/ScAlN QWs as an exceptionally strong second-order nonlinear platform in the near-infrared, with effective sheet nonlinearities far exceeding those of bulk materials commonly employed in SPDC sources.

Using this framework, metasurface geometries were inverse designed that simultaneously optimize incoupling of a normally incident pump, field confinement in the MQW region, and radiative outcoupling of the down-converted photons.  The calculated nonlinear overlap and quality factors yield biphoton generation rates on the order of $10^{10}$~s$^{-1}$ per unit cell for 1~W of incident pump power at telecom wavelengths, in an overall thickness of only $\sim 0.5~\mu$m.  
This performance is comparable to that of state-of-the-art mode-matched high-$Q$ microresonators based on bulk III–V nonlinearities, while offering a dramatically reduced footprint, relaxed phasematching constraints, and efficient free-space or fiber coupling at both pump and signal/idler frequencies.  
More broadly, the theoretical and design tools developed here provide a general route to quantitatively predict and engineer SPDC in open, lossy nanophotonic systems where modal overlap, radiative losses, and material absorption are all comparable and must be treated on equal footing.  
The same framework can be extended to include electrically tunable bandstructures, more complex multi-resonant or multi-QW designs, and other nonlinear processes such as sum- and difference-frequency generation in the near- and mid-infrared.  
On the application side, integrating such metasurface-based entangled photon sources with existing III–nitride and silicon photonics platforms opens a realistic pathway toward compact, monolithically integrated quantum light sources for fiber-based quantum communication, on-chip quantum information processing, and quantum sensing in the technologically important telecom window.

\section{Acknowledgments}
J.Y., S. A, I. B, R.S. acknowledge support from the U.S. Department of Energy, Office of Basic Energy Sciences, Division of Materials Sciences and Engineering. A.B acknowledges support from the Laboratory Directed Research and Development program at Sandia National Laboratories and Sandia University Partnerships Network (SUPN) program. This work was performed in part at the Center for Integrated Nanotechnologies, an Office of Science User Facility operated for the U.S. Department of Energy (DOE) Office of Science. Sandia National Laboratories is a multimission laboratory managed and operated by National Technology \& Engineering Solutions of Sandia, LLC, a wholly owned subsidiary of Honeywell International, Inc., for the U.S. DOE’s National Nuclear Security Administration under Contract No. DE-NA-0003525. The views expressed in the article do not necessarily represent the views of the U.S. DOE or the United States Government.

\section{Conflicts of Interest}
The authors declare no conflicts of interest.

\section{Data availability}

The data that support the findings of this study are available from the corresponding authors upon reasonable request.

\bibliography{refs_main}

\end{document}


\title{Efficient generation of entangled photons in the telecommunications range using nonlinear metasurfaces integrated with ScAlN/GaN heterostructures.\\ Supporting information}

\author{Jaeyeon Yu}
\affiliation{Center for Integrated Nanotechnologies, Sandia National Laboratories, Albuquerque, New Mexico, USA}

\author{Jewel Mohajan}
\affiliation{Princeton University, Princeton, NJ, USA}

\author{Mikhail Tokman}
\affiliation{Ariel University, Ariel, Israel}

\author{Jackson Stewart}
\affiliation{Texas A\&M University, College Station, TX, 77843 USA}

\author{Anthony Rice}
\affiliation{Sandia National Laboratories, Albuquerque, New Mexico, USA}

\author{Sadhvikas Addamane}
\affiliation{Center for Integrated Nanotechnologies, Sandia National Laboratories, Albuquerque, New Mexico, USA}

\author{Oana Malis}
\affiliation{Purdue University, West Lafayette, IN, USA}

\author{Alejandro W. Rodriguez}
\affiliation{Princeton University, Princeton, NJ, USA}

\author{Igal Brener}
\affiliation{Center for Integrated Nanotechnologies, Sandia National Laboratories, Albuquerque, New Mexico, USA}

\author{Raktim Sarma*}
\affiliation{Center for Integrated Nanotechnologies, Sandia National Laboratories, Albuquerque, New Mexico, USA}

\author{Alexey Belyanin}
\affiliation{Texas A\&M University, College Station, TX, 77843 USA}

\date{\today}

\begin{abstract}
    This document gives the parameters of the quantum well heterostructures used in modeling, presents the details of the solution of Heisenberg-Langevin equations for signal and idler fields, and explores the formulation of the inverse design problem for down conversion enhancement taking advantage of the gradient based adjoint methods for topology optimization of metasurfaces integrated with GaN/ScAlN QWs that support intersubband transitions with nonlinear responses. 
\end{abstract}
\maketitle 

\noindent
\textbf{\#Corresponding author:} Raktim Sarma (rsarma@sandia.gov)\\
\noindent
\textbf{Equal contribution:} Jaeyeon Yu and Jewel Mohajan contributed equally to this work.

\section{Parameters of ISB transitions in GaN quantum wells}

The unit-period heterostructures consisting of ScAlN/AlN/GaN were optimized, and the ISB energies and dipole matrix elements were calculated using the \textit{nextnano} software. The electronic subband energies and wavefunctions were obtained by solving the self-consistent Poisson-Schr\"odinger equation, and the ISB optical matrix elements were evaluated from the resulting envelope functions. The relevant dipole matrix elements were defined as $z_{ij}=\langle i|z|j\rangle$, where $z$ denotes the growth-direction coordinate (surface-normal direction). The diagonal (permanent) elements $z_{ii}=\langle i|z|i\rangle$ corresponding to the position expectation values were also extracted.

For the single quantum well (Fig.~2a), the calculated transition energies were $E_{12}=0.790$~eV, $E_{13}=1.382$~eV, and $E_{14}=1.605$~eV. The corresponding dipole matrix elements were $z_{12}=0.382$~nm, $z_{23}=0.323$~nm, $z_{13}=0.033$~nm, $z_{24}=0.348$~nm, $z_{34}=0.797$~nm, and $z_{14}=0.035$~nm. The diagonal position expectation values were $z_{11}=4.30$~nm, $z_{22}=4.05$~nm, $z_{33}=5.83$~nm, and $z_{44}=3.62$~nm.

For the coupled double-quantum-well structure (Fig.~2b), the calculated transition energies were $E_{12}=0.768$~eV, $E_{13}=0.799$~eV, $E_{14}=1.398$~eV, and $E_{15}=1.602$~eV. The dipole matrix elements were $z_{12}=0.259$~nm, $z_{13}=0.286$~nm, $z_{24}=0.137$~nm, $z_{25}=0.112$~nm, $z_{34}=0.316$~nm, $z_{35}=0.348$~nm, $z_{14}=0.036$~nm, and $z_{15}=0.034$~nm. The diagonal position expectation values were $z_{11}=6.00$~nm, $z_{22}=4.41$~nm, $z_{33}=4.78$~nm, $z_{44}=7.45$~nm, and $z_{55}=6.45$~nm.

In both structures, the ISB transitions were designed to be energetically nearly equidistant, such that the corresponding resonant wavelengths were placed near 1.55~$\mu$m. The second-order nonlinear susceptibility associated with ISB transitions was analyzed by separating the contributions into (i) resonant terms proportional to products of three off-diagonal dipole matrix elements (for example, $z_{12}z_{23}z_{31}$) and (ii) nonresonant terms involving a product of an off-diagonal dipole matrix element and a diagonal dipole difference (for example, $|z_{12}|^2 \Delta z$ with $\Delta z=z_{22}-z_{11}$). Comparable magnitudes were obtained for the resonant contributions in the single and coupled double-quantum-well designs. In contrast, a stronger diagonal nonresonant contribution was obtained in the coupled double-quantum-well structure, consistent with the enhanced asymmetry quantified by the diagonal expectation values.

\section{Solution of Heisenberg-Langevin equations for quantized signal and idler fields}

The solution to Eqs.~(22) in the main text can be represented as~\cite{et2017,belyanin2013}
\begin{equation}
\label{Eq:69}
	\left(\begin{array}{c}\hat{c}_{0s}\\\hat{c}_{0i}^\dagger\end{array}\right)
=\left(\begin{array}{c}1\\K_1\end{array}\right)e^{-\lambda_1t}
\left(\hat{c}_1+\int\limits_0^te^{\lambda_1t'}\hat{L}_1(t')dt' \right)
+\left(\begin{array}{c}1\\K_2\end{array}\right)e^{-\lambda_2t}
\left(\hat{c}_2+\int\limits_0^te^{\lambda_2t'} \hat{L}_2(t')dt' \right),
\end{equation}
where $\lambda_{1,2}$ and $\left(\begin{array}{c}1\\K_{1,2}\end{array}\right)$ are eigenvalues and eigenvectors of the $2\times2$ matrix:
\begin{eqnarray}
\label{Eq:70}
	\left(\begin{array}{cc}
\Gamma_s& g_s\\ g_i^*&\Gamma_i
\end{array}\right)
\times\left(\begin{array}{c}1\\K_{1,2}\end{array}\right)
=\lambda_{1,2}\left(\begin{array}{c}1\\K_{1,2}\end{array}\right),
\\
\label{Eq:71}
	\left.\begin{array}{cc}
\hat{c}_1=\displaystyle\frac{K_2\hat{c}_s(0)-\hat{c}_i^\dagger(0)}{K_2-K_1},&
\hat{c}_2= - \displaystyle\frac{K_1\hat{c}_s(0)-\hat{c}_i^\dagger(0)}{K_2-K_1}\\
\hat{L}_1=\displaystyle\frac{K_2\hat{L}_s-\hat{L}_i^\dagger}{K_2-K_1},&
\hat{L}_2= - \displaystyle\frac{K_1\hat{L}_s-\hat{L}_i^\dagger}{K_2-K_1}
\end{array}\right\},
\end{eqnarray}
$\hat{c}_s(0)$ and $\hat{c}_i^\dagger(0)$ are initial conditions.

The solution for eigenvalues  is 
\begin{equation}
\label{eigen}
	 \lambda_{1,2} = \frac{\Gamma_s + \Gamma_i}{2} \pm \sqrt{ g_s g_i^* - \left( \frac{\Gamma_s + \Gamma_i}{2}\right)^2}. 
\end{equation}
The parametric instability occurs if the real part of one of the eigenvalues is greater than zero. There is a plethora of interesting dynamic regimes beyond parametric down-conversion, depending on the phase of the product $g_s g_i^*$. For example, when all fields are resonant with corresponding intersubband transitions, it is the strong {\it low-frequency} driving field, say at the signal frequency, which gives rise to generation of weak high-frequency ($\omega_p$) and low-frequency ($\omega_i$)  fields. These regimes have been studied in connection with various lasing without inversion schemes; see, e.g., \cite{belyanin2001a}. Since we are interested in maximizing the parametric down-conversion process of the decay of a strong high-frequency pump, it is beneficial to have the fields detuned from the corresponding ISB resonances by a couple of linewidths to reduce optical pumping, resonant absorption, and uncorrelated spontaneous emission at the ISB transitions. In this case one can neglect the imaginary part of $\chi^{(2)}$  and assume that  $g = \sqrt{g_s g_i^*}$ is real and positive. In fact, in this regime $g_s \approx g_i$. 
Then we recover the familiar condition for the parametric instability, 
\begin{equation}
\label{Eq:72}
	g^2 = g_s g_i^* > \Gamma_s\Gamma_i.
\end{equation}

To avoid cumbersome expressions, consider the decay of a pump photon into identical quanta. Then Eqs.~(\ref{Eq:69})-(\ref{Eq:71}) yield
\begin{multline}
\label{Eq:75}
	\hat{c}_{0s}=e^{-\Gamma_st}\left[\hat{c}_s(0)\text{cosh}( g t)-\hat{c}_s^\dagger(0)\text{sinh}( g t)\right]
+\int\limits_0^te^{(- g+\Gamma_s)(t'-t)}\frac{\hat{L}_s(t')-\hat{L}_s^\dagger(t')}{2}dt'+\\
+\int\limits_0^te^{( g+\Gamma_s)(t'-t)}\frac{\hat{L}_s(t')+\hat{L}_s^\dagger(t')}{2}dt'.
\end{multline}
Taking into account the properties of Langevin operators in Eq.~(22) and taking $	\left<\hat{c}_s^\dagger(0)\hat{c}_s^\dagger(0)\right>=\left<\hat{c}_s(0)\hat{c}_s(0)\right>=0$ as an initial state, one can derive from Eq.~(\ref{Eq:75}) the average photon numbers for signal modes $	n_s=\left< \hat{c}_{s}^\dagger\hat{c}_{s}\right>=\left<\hat{c}_{0s}^\dagger\hat{c}_{0s}\right>$:
\begin{multline}
\label{Eq:76}
	n_s=e^{-2\Gamma_st}\left\{n_s(0)\left[ \text{cosh}^2( g t)+\text{sinh}^2( g t)\right]+\text{sinh}^2( g t) \right\}\\
+\left[\Gamma_{\sigma s}n_{T_\sigma}(\omega_s)+\Gamma_{rs}n_{T_r}(\omega_s)\right]
\times\left(\frac{1-e^{2( g-\Gamma_s)t}}{2(- g+\Gamma_s)}+\frac{1-e^{2(- g-\Gamma_s)t}}{2( g+\Gamma_s)}\right)\\
+\Gamma_s
\left(\frac{1-e^{2( g-\Gamma_s)t}}{4(- g+\Gamma_s)}+\frac{1-e^{2(- g-\Gamma_s)t}}{4( g+\Gamma_s)}-\frac{1-e^{-2\Gamma_st}}{2\Gamma_s}\right),
\end{multline}
where $\Gamma_s=\Gamma_{\sigma s}+\Gamma_{rs}$. When the parametric amplification starts from the level of vacuum fluctuations, one should put $n_s(0)=0$ in Eq.~(\ref{Eq:76}). 

In the limit of zero pumping intensity, Eq.~(\ref{Eq:76}) gives an expression which describes how the initial perturbation of a photon number approaches equilibrium: 
\begin{equation}
\label{Eq:77}
	n_s=e^{-2\Gamma_st}n_s(0)+\frac{\Gamma_{\sigma s}n_{T_\sigma}(\omega_s)+\Gamma_{rs}n_{T_r}(\omega_s)}{\Gamma_s}\times\left(1-e^{-2\Gamma_st}\right).
\end{equation}

Above the instability threshold, when $ g\gg\Gamma_s$, it is enough to keep only exponentially growing terms in Eq.~(\ref{Eq:76}). We can also assume that an average number of thermal photons in an ambient space $n_{T_r}(\omega_s)$ is negligible. This gives an expression for the parametric signal power $P_s	\approx 2 g\hbar\omega_s n_s$:
\begin{equation}
\label{Eq:78}
	P_s\approx  g\hbar\omega_s e^{2 g t}
\left[ n_s(0)+\frac{\Gamma_{\sigma s}}{ g}n_{T_\sigma}(\omega_s)+\frac{1}{2} \right].
\end{equation}
Obviously this expression is valid only at the initial linear stage. The subsequent evolution is governed by the nonlinear pump depletion and nonlinear modification of phasematching conditions and losses.  An order-of magnitude estimate of the maximum steady-state power  can be obtained from Manley-Rowe relations as shown below for a specific example. 

The fractions of the power emitted outside and absorbed inside a cavity are $P_{r s}\approx\Gamma_{r s}P_s/ g$ and $P_{\sigma s}\approx\Gamma_{\sigma s}P_s/ g$ respectively; most of the power is accumulated in a cavity. From Eq.~(\ref{Eq:78}) one can see that the amplification of intrinsic thermal noise of a QW layer with temperature $T_\sigma$ can be ignored if $\displaystyle\frac{\Gamma_{\sigma s}}{ g}\cdot\frac{2}{\text{exp}(\hbar\omega_s/T_\sigma)-1}\ll1$. 

The limit of the SPDC corresponds to $g \ll \Gamma_s$. 
In the stationary limit, when $(\Gamma_s -  g) t \rightarrow \infty$, the radiated signal power  $P_{rs} \approx 2 \Gamma_{rs} \hbar \omega_s n_s$ is given by Eq.~(25) in the main text.

\section{Details of the Inverse Design Formalism}
For the optimization, we used NLopt library which is an open-source library for nonlinear optimization. We tried several global optimization algorithms that are available within the library such as Controlled Random Search (CRS) with local mutation, Dividing Rectangles (DIRECT) algorithm, etc. However, for highly non-convex problems with thousands of degrees of freedom such as the one presented in this work, the gradient-independent global optimization methods are very slow to converge, and without prior knowledge of upper bounds on objectives it is often impossible to decide when to stop the optimization process. On the other hand, the adjoint-based gradient dependent optimization allows us to calculate the gradient very efficiently and guarantees rapid convergence to a local optimum that depends primarily on our choice of initial device structure, which is taken care of by ensuring a high nonlinear overlap between modes at both the fundamental and the second harmonic frequency in our first optimization step. We therefore chose the adjoint-based gradient-dependent optimization. We found that the globally-convergent method-of-moving-asymptotes (MMA) algorithm for gradient-based local optimization is the most effective one with fast convergence, and reported the best performing devices within this framework~\cite{NLopt}.

Finally, in our optimization framework, the degrees of freedom (device pixels) are initially allowed to assume any value between 0 and 1 (grayscale) before applying binarization filters in the final step. During the filtering process we also implement blurring operators of different radii to smooth out the device to account for fabrication resolution limits and/or minimum feature size. Both the blurring and filtering operations are performed on the fly with the optimization process so that we remain in the vicinity of the optimum. In principle, such blurring operations can be adopted to integrate probabilistic models or stochastic effects to mitigate fabrication errors given that we can estimate the imperfections encountered during the manufacturing process. 

In the following sections, we describe the mathematical formulation for inverse design~\cite{molesky2018inverse,Lin_Optica_2016,Stich_ACSnano_2025,Mohajan_OE_2023}:

\section{Problem set up for maximizing SHG}
\subsection{Maxwell's Equation in Frequency Domain}
Let us define the fields as time harmonic and consider a single frequency $\omega$ 
\begin{align}
    \vb{E}(\vb{r},t)&=\vb{E}(\vb{r}) \;e^{i\omega t} \\
    \vb{H}(\vb{r},t)&=\vb{H}(\vb{r}) \;e^{i\omega t}
\end{align}
Consider $\vb{E}(\vb{r})$ to be complex numbers but in the end all the fields as well as all the calculated quantity must be real.\\
\\
Real fields:
\begin{align}
    \vb{E}&=\re{\vb{E}(\vb{r}) \;e^{i\omega t}} = \frac{1}{2}\big[ \vb{E}(\vb{r})\;e^{i\omega t} + \vb{E}^*(\vb{r})\;e^{-i\omega t} \big] \\
    \vb{H}&=\re{\vb{H}(\vb{r}) \;e^{i\omega t}} = \frac{1}{2}\big[ \vb{H}(\vb{r})\;e^{i\omega t} + \vb{H}^*(\vb{r})\;e^{-i\omega t} \big]
\end{align}
In this section, we start from Maxwell's equation and define Maxwell's operator which is used to formulate the objective and gradients required for the topology optimization~\cite{molesky2018inverse,Mohajan_OE_2023,Lin_Optica_2016}.
From Maxwell's equations:
\begin{align}
    \curl{\vb{H}}(\vb{r},t) &= \vb{J}(\vb{r},t) + \frac{\partial \vb{D}(\vb{r},t)}{\partial t} = \vb{J}(\vb{r},t) + \epsilon(\vb{r}) \frac{\partial \vb{E}(\vb{r},t)}{\partial t}  
    \\
    \curl{\vb{H}}(\vb{r})e^{iwt} &= \vb{J}(\vb{r})e^{iwt}  + iw\epsilon(\vb{r})\vb{E}(\vb{r})e^{iwt}
    \\
    \curl{\vb{H}}(\vb{r}) &= \vb{J}(\vb{r}) + iw\epsilon(\vb{r})\vb{E}(\vb{r})
\end{align}
\\
Faraday's law:
\begin{align}
    \curl{\vb{E}}(\vb{r},t) &= -\frac{\partial \vb{B}(\vb{r},t)}{\partial t} 
    \\
    \curl{\vb{E}}(\vb{r})e^{iwt} &= -\mu\frac{\partial \vb{H}(\vb{r})e^{iwt}}{\partial t} 
    \\
    \curl{\vb{E}}(\vb{r}) &= -iw\mu\vb{H}(\vb{r},t)
\end{align}
\\
Taking curl on both side and setting $\mu=1$(non-magnetic material):
\begin{align}
    \curl{\curl{\vb{E}}}(\vb{r}) &= -iw\curl{\vb{H}}(\vb{r})
    \\
    \curl{\curl{\vb{E}}}(\vb{r}) &= -iw\vb{J}(\vb{r}) + w^2\epsilon(\vb{r})\vb{E}(\vb{r})
    \\
    \curl{\curl{\vb{E}}}(\vb{r}) - w^2\epsilon(\vb{r})\vb{E}(\vb{r}) &= -iw\vb{J}(\vb{r})
    \\
    (\curl\curl - w^2\epsilon(\vb{r}))\vb{E}(\vb{r}) &= -iw\vb{J}(\vb{r})
    \\
    \mathbb{M}\vb{E} &= -iw\vb{J}
    \\
    \vb{E} &= \mathbb{M}^{-1}\big(-iw\vb{J}\big)
\end{align}
\\
where we defined Maxwell Operator, $\mathbb{M}=\curl\curl - \epsilon(\vb{r}) w^2$ and $\epsilon(\vb{r}) = \epsilon_{bg} + \rho(\vb{r})\left(\epsilon_{design} - \epsilon_{bg}\right)$\\ where $\epsilon_{bg}$ and $\epsilon_{design}$ are the dielectric constants of the background(vacuum) and of the nonlinear metasurface.
\\ \\
For each pixel(voxel) in the design domain, let's define for the i'th pixel
\begin{align}
    \epsilon_i &= \epsilon_{bg} + \rho_i\Delta\epsilon
    \\
    \Delta\epsilon &= \epsilon_{design} - \epsilon_{bg}.
\end{align}
Gradients of Maxwell's Operator $\mathbb{M}$, and hence, of the inverse Maxwell operator $\mathbb{M}^{-1}$ are calculated as
\begin{align}
    \frac{\partial\mathbb{M}}{\partial \rho_i} &= \frac{\partial}{\partial \rho_i}(\curl\curl - \epsilon_i w^2) = -\delta_{ii}\Delta\epsilon w^2
    \\
    \frac{\partial \mathbb{M}^{-1}}{\partial \rho_i} &= -\mathbb{M}^{-1} \frac{\partial \mathbb{M}}{\partial \rho_i} \mathbb{M}^{-1} = \mathbb{M}^{-1}\bigg(\delta_{ii}\Delta\epsilon w^2 \bigg)\mathbb{M}^{-1}
\end{align}
where $\delta_{ii}$ is a projection operator that projects onto the specific design voxel.
\subsection{Second harmonic analysis: nonnlinear material($\chi^{(2)}$)}
For a material with nonlinear response we can calculate the fields both at the fundamental frequency and at the second harmonic as
\begin{align}
    \vb{E}_1 &= \mathbb{M}_1^{-1}(-iw_1\vb{J}_1) \; \; \; \text{with,} \; \mathbb{M}_1=\curl\curl\; -\;\epsilon_1(\vb{r})\omega_1^2
    \\
    \vb{E}_2 &= \mathbb{M}_2^{-1}(-iw_2\vb{J}_2) \; \; \; \text{with,} \; \mathbb{M}_2=\curl\curl\; -\;\epsilon_2(\vb{r})\omega_2^2
\end{align}
where $\vb{J}_1$ is the initial incident source. We used non-depleted pump approximation and weak nonliearity assumption to decouple the equations at $w_1$ and $w_2$.
Second harmonic polarization, $\vb{P}_2(\vb{r},t) =\epsilon_0\chi^{(2)}(\vb{r})\vb{E}_{1}^2 e^{iw_2t}$,  and extended current $\vb{J}_{2}(\vb{r}) = \frac{\partial \vb{P}_{2}(\vb{r},t)}{\partial t} = iw_2\epsilon_0\chi^{(2)}(\vb{r})\vb{E}_{1}^2(\vb{r})e^{iw_2t}$\\
For the all-dielectric metasurface with only non-zero element of $\chi^{(2)}$ being  $\chi_{zzz}^{(2)}$,
\begin{align}
    {P}_{2z}(\vb{r},t) &=\epsilon_0\chi_{zzz}^{(2)}(\vb{r}){E}_{1z}^2(\vb{r}) \; e^{iw_2t}
    \\
    {P}_{2x}(\vb{r}) &={P}_{2y}(\vb{r})=0
    \\
    {J}_{2z}(\vb{r},t) &= \frac{\partial {P}_{2z}(\vb{r},t)}{\partial t} = iw_2\epsilon_0\chi_{zzz}^{(2)}(\vb{r}){E}_{1z}^2(\vb{r})e^{iw_2t} 
    \\ 
    \text{where,} \; w_2&=2w_1; 
    \nonumber
    \\
    {J}_{2x}(\vb{r}) &= {J}_{2y}(\vb{r}) = 0
\end{align}

\section{Maximizing overlap - optimizing for total extracted power from the extended source}

For the auxiliary optimization problem,
Objective:
\begin{align}
    \mathbb{O}=-\frac{1}{2}\re{\int_{\Omega}\vb{J}_2^*.\vb{E}_2\;d\vb{r}}
\end{align}
where $\Omega$ is the whole computational domain.
\\
Gradient of the objective with respect to the design degrees of freedom:
\begin{align}
    \frac{\partial\mathbb{O}}{\partial \rho_i}
    = -\frac{1}{2}\re{\int_{\Omega}\frac{\partial}{\partial \rho_i}\bigg(\vb{J}_2^{*T}.\vb{E}_2\bigg)\;d\vb{r}}
\end{align}
Let's look at the term inside the integral
\begin{align}
    \frac{\partial}{\partial \rho_i}\bigg(\vb{J}_2^{*T}.\vb{E}_2\bigg)
    = \bigg(\frac{\partial\vb{J}_2}{\partial \rho_i}\bigg)^{*T}\vb{E}_2
    + \vb{J}_2^{*T}\frac{\partial\vb{E}_2}{\partial \rho_i}
\end{align}
We can calculate each term separately, and plug them back in to get the gradient
\begin{align}
    \frac{\partial\vb{J}_2}{\partial \rho_i}
    &= \frac{\partial}{\partial \rho_i}\bigg(iw_2\epsilon_0\chi^{(2)}\vb{E}_1^2 \bigg)
    \\
    &= \delta_{ii} iw_2\epsilon_0\chi^{(2)}\vb{E}_1^2 
    + iw_2\epsilon_0\chi^{(2)} 2\vb{E}_1\frac{\partial \vb{E}_1}{\partial \rho_i}
    \\
    &= \delta_{ii} iw_2\epsilon_0\chi^{(2)}\vb{E}_1^2 
    + iw_2\epsilon_0\chi^{(2)} 2\vb{E}_1\mathbb{M}_1^{-1}(\delta_{ii}\Delta\epsilon_1 w_1^2)\vb{E}_1
    \\
    \frac{\partial \vb{E}_2}{\partial \rho_i}
    &= \mathbb{M}_2^{-1} \big( \delta_{ii}\Delta\epsilon_2 w_2^2\big) \vb{E}_2 
    + \mathbb{M}_2^{-1} (-iw_2) \delta_{ii} iw_2\epsilon_0\chi^{(2)}\vb{E}_{1}^2 
    \nonumber\\
    &+ \mathbb{M}_2^{-1} (-iw_2) iw_2\epsilon_0\chi^{(2)} 2\vb{E}_{1} \mathbb{M}_1^{-1} \big(\delta_{ii}\Delta\epsilon_1\omega_1^2\big) \vb{E}_1
\end{align}
Now,
\begin{align}
   \bigg(\frac{\partial\vb{J}_2}{\partial \rho_i} \bigg)^{*T} \vb{E}_2
   &= \bigg(\delta_{ii} iw_2\epsilon_0\chi^{(2)}\vb{E}_1^2\bigg)^{*T} \vb{E}_2 
    + \bigg(iw_2\epsilon_0\chi^{(2)} 2\vb{E}_1\mathbb{M}_1^{-1}(\delta_{ii}\Delta\epsilon_1 w_1^2)\vb{E}_1\bigg)^{*T} \vb{E}_2
    \\
    &= \bigg(\big(\delta_{ii} iw_2\epsilon_0\chi^{(2)}\vb{E}_1^2\big)^{T} \vb{E}_2^*\bigg)^*
    + \vb{E}_1^{*T}(\delta_{ii}\Delta\epsilon_1w_1^2)^* \mathbb{M}_1^{-1^*}\big(iw_2\epsilon_0\chi^{(2)} 2\vb{E}_1\big)^{*T}\vb{E}_2
    \\
    &= \bigg(\big(\delta_{ii} iw_2\epsilon_0\chi^{(2)}\vb{E}_1^2\big)^{T} \vb{E}_2^*\bigg)^*
    + \bigg(\vb{E}_1^{T}(\delta_{ii}\Delta\epsilon_1w_1^2) \mathbb{M}_1^{-1}\big(iw_2\epsilon_0\chi^{(2)} 2\vb{E}_1\big)^{T}\vb{E}_2^*\bigg)^*
    \\
    &= \bigg(\big(\delta_{ii} iw_2\epsilon_0\chi^{(2)}\vb{E}_1^2\big)^{T} \vb{E}_2^*\bigg)^*
    + \bigg(\vb{E}_1^{T}(\delta_{ii}\Delta\epsilon_1w_1^2) \vb{u}_2\bigg)^* 
    \\
    &\text{where,\;}\vb{u}_2=\mathbb{M}_1^{-1}\bigg(\big(iw_2\epsilon_0\chi^{(2)} 2\vb{E}_1\big)^{T}\vb{E}_2^*\bigg)
\end{align}
And
\begin{align}
    \vb{J}_2^{*T}\frac{\partial\vb{E}_2}{\partial \rho_i}
    &= \vb{J}_2^{*T}
    \mathbb{M}_2^{-1} \big( \delta_{ii}\Delta\epsilon_2 w_2^2\big) \vb{E}_2 
    + \vb{J}_2^{*T}
    \mathbb{M}_2^{-1} (-iw_2) \delta_{ii} iw_2\epsilon_0\chi^{(2)}\vb{E}_{1}^2 
    \nonumber\\
    &+ \vb{J}_2^{*T}
    \mathbb{M}_2^{-1} (-iw_2) iw_2\epsilon_0\chi^{(2)} 2\vb{E}_{1} \mathbb{M}_1^{-1} \big(\delta_{ii}\Delta\epsilon_1\omega_1^2\big) \vb{E}_1
    \\
    &= \bigg(\mathbb{M}_2^{-1}(-iw_2\vb{J}_2^{*})\bigg)^T
     \big( \delta_{ii}\Delta\epsilon_2 iw_2\big) \vb{E}_2 
    + \bigg(\mathbb{M}_2^{-1} (-iw_2\vb{J}_2^{*})\bigg)^T
     \delta_{ii} iw_2\epsilon_0\chi^{(2)}\vb{E}_{1}^2 
    \nonumber\\
    &+ \bigg(\mathbb{M}_2^{-1} (-iw_2\vb{J}_2^{*})\bigg)^T
     iw_2\epsilon_0\chi^{(2)} 2\vb{E}_{1} \mathbb{M}_1^{-1} \big(\delta_{ii}\Delta\epsilon_1\omega_1^2\big) \vb{E}_1
     \\
     &= \vb{u}_1^T
     \big( \delta_{ii}\Delta\epsilon_2 iw_2\big) \vb{E}_2 
    + \vb{u}_1^T
     \delta_{ii} iw_2\epsilon_0\chi^{(2)}\vb{E}_{1}^2 
    + \vb{u}_1^T
     iw_2\epsilon_0\chi^{(2)} 2\vb{E}_{1} \mathbb{M}_1^{-1} \big(\delta_{ii}\Delta\epsilon_1\omega_1^2\big) \vb{E}_1
     \\
     &\text{where,\;}\vb{u}_1=\mathbb{M}_2^{-1} (-iw_2\vb{J}_2^{*})
     \nonumber\\
     &= \vb{u}_1^T
     \big( \delta_{ii}\Delta\epsilon_2 iw_2\big) \vb{E}_2 
    + \vb{u}_1^T
     \delta_{ii} iw_2\epsilon_0\chi^{(2)}\vb{E}_{1}^2 
    + \bigg(\mathbb{M}_1^{-1}\big(iw_2\epsilon_0\chi^{(2)} 2\vb{E}_{1}\big)^T\vb{u}_1\bigg)^T
    \big(\delta_{ii}\Delta\epsilon_1\omega_1^2\big) \vb{E}_1
    \\
    &= \vb{u}_1^T
     \big( \delta_{ii}\Delta\epsilon_2 iw_2\big) \vb{E}_2 
    + \vb{u}_1^T
     \delta_{ii} iw_2\epsilon_0\chi^{(2)}\vb{E}_{1}^2 
    + \vb{u}_3^T
    \big(\delta_{ii}\Delta\epsilon_1\omega_1^2\big) \vb{E}_1
    \\
    &\text{where,\;}\vb{u}_3=\mathbb{M}_1^{-1}\bigg(\big(iw_2\epsilon_0\chi^{(2)} 2\vb{E}_{1}\big)^T\vb{u}_1\bigg).
    \nonumber
\end{align}

\section{Maximizing z-polarized Electric Field inside Nitride Quantum-Well - The coupling problem}
Consider the following optimization problem for maximizing coupling at both pump frequency $\omega_p$ and signal frequency $\omega_s$:
\begin{equation}
\begin{aligned}
\max_{\overline{\rho}} \quad &  \Phi_{p,s} =  \int_{QW} \abs{E_{z_{p,s}}}^2 \ d\vb{r}  
\end{aligned}
\end{equation}
Objective:
\begin{align}
    \mathbb{O} = \int_{QW}\frac{\partial}{\partial \rho_i}\bigg((\mathbb{P}\vb{E}_{p,s})^{*T}.(\mathbb{P}\vb{E}_{p,s})\bigg)\;d\vb{r},
\end{align}
where $\mathbb{P}$ is the operator that picks out the z-component of the electric field.

Now, gradient of the objective:
\begin{align}
    \frac{\partial}{\partial \rho_i} \bigg( (\mathbb{P}\vb{E})^{*T}(\mathbb{P}\vb{E}) \bigg)
    &= \big(\mathbb{P}\frac{\partial \vb{E}}{\partial \rho_i} \big)^{*T}\mathbb{P}\vb{E}
    + \big(\mathbb{P}\vb{E} \big)^{*T}\big(\mathbb{P}\frac{\partial \vb{E}}{\partial \rho_i}\big)
    \\
    &= \big( \frac{\partial \vb{E}}{\partial \rho_i} \big)^{*T}\mathbb{P}^T\mathbb{P}\vb{E} 
    + \vb{E}^{*T}\mathbb{P}^T\mathbb{P}\frac{\partial \vb{E}}{\partial \rho_i}
    \\
    &= \big( \frac{\partial \vb{E}}{\partial \rho_i} \big)^{*T}\mathbb{P}\vb{E} 
    + \vb{E}^{*T}\mathbb{P}\frac{\partial \vb{E}}{\partial \rho_i}
    \\
    &= \big( \frac{\partial \vb{E}}{\partial \rho_i} \big)^{*T}\mathbb{P}\vb{E} 
    + \big(\frac{\partial \vb{E}}{\partial \rho_i}\big)^{T}\mathbb{P}\vb{E}^* & \text{Scalar;\;} c^T=c
    \\
    &= \big( \frac{\partial \vb{E}}{\partial \rho_i} \big)^{*T}\mathbb{P}\vb{E} 
    + \bigg(\big(\frac{\partial \vb{E}}{\partial \rho_i}\big)^{*T}\mathbb{P}\vb{E}\bigg)^*
    \\
    &= 2\re{\big( \frac{\partial \vb{E}}{\partial \rho_i} \big)^{*T}\mathbb{P}\vb{E}}
    \\
    &= 2\re{\big( \frac{\partial} {\partial \rho_i}\mathbb{M}^{-1}(-iw\vb{J}) \big)^{*T}\mathbb{P}\vb{E}}
    \\
    &= 2\re{\big( \mathbb{M}^{-1}(\delta_{ii}\Delta\epsilon.w^2)\mathbb{M}^{-1}(-iw\vb{J}) \big)^{*T}\mathbb{P}\vb{E}}
    \\
    &= 2\re{\big( \mathbb{M}^{-1}(\delta_{ii}\Delta\epsilon.w^2)\vb{E} \big)^{*T}\mathbb{P}\vb{E}}
    \\
    &= 2\re{\vb{E}^{*T}(\delta_{ii}\Delta\epsilon.w^2)^*\mathbb{M}^{-1^*}\mathbb{P}\vb{E}}
    \\
    &= 2\re{\vb{E}^{T}(\delta_{ii}\Delta\epsilon.w^2)\mathbb{M}^{-1}\mathbb{P}\vb{E}^*}^*
    \\
    &= 2\re{\vb{E}^{T}(\delta_{ii}\Delta\epsilon.w^2)\mathbb{M}^{-1}\mathbb{P}\vb{E}^*} & \re{z^*}=\re{z}
    \\
    \frac{\partial}{\partial \rho_i} \bigg( (\mathbb{P}\vb{E})^{*T}(\mathbb{P}\vb{E}) \bigg)
    &= 2\re{\vb{E}^{T}(\delta_{ii}\Delta\epsilon.w^2)\vb{u}} & \vb{u}=\mathbb{M}^{-1}\big(\mathbb{P}\vb{E}^* \big)
\end{align}

\bibliography{refs_supp}